\documentclass{aa}  
\usepackage{graphicx}
\usepackage{txfonts}
\usepackage{nccmath}
\usepackage{xcolor}
 
\begin{document} 

    \title{XXL-HSC: An updated catalogue of high-redshift ($z\geqslant3.5$)\\ X-ray AGN in the XMM-XXL northern field\thanks{Based on observations obtained with XMM-Newton, an ESA science mission with instruments and contributions directly funded by ESA Member States and NASA.}}
    \subtitle{Constraints on the bright end of the soft logN-logS}

    \titlerunning{High-z X-ray AGN}
    \authorrunning{E. Pouliasis et al.}

    \author{E.~Pouliasis\inst{1}
    \and I.~Georgantopoulos\inst{1}
    \and A.~Ruiz\inst{1}
    \and R.~Gilli\inst{2}
    \and E.~Koulouridis\inst{1}
    \and M.~Akiyama\inst{3}
    \and  Y.~Ueda\inst{4}
    \and L.~Chiappetti\inst{5}	
    \and C.~Garrel\inst{6}
    \and C.~Horellou\inst{7}
    \and T.~Nagao\inst{8}
    \and S.~Paltani\inst{9} 	
    \and M.~Pierre\inst{6}
    \and Y.~Toba\inst{4,8,9}
    \and C.~Vignali\inst{11,2}}

    \institute{IAASARS, National Observatory of Athens, Ioannou Metaxa and Vasileos Pavlou GR-15236, Athens, Greece\\
    \email{epouliasis@noa.gr}
    \and
    INAF - Osservatorio di Astrofisica e Scienza dello Spazio di Bologna, Via Gobetti 93/3, I-40129 Bologna, Italy
    \and
    Astronomical Institute, Tohoku University, 6-3 Aramaki, Aoba-ku, Senda, 980-8578, Japan
    \and
    Department of Astronomy, Kyoto University, Kitashirakawa-Oiwake-cho, Sakyo-ku, Kyoto 606-8502, Japan
    \and
    INAF, IASF Milano, via Corti 12, Milano, I-20133, Italy
    \and 
    AIM, CEA, CNRS, Université Paris-Saclay, Université Paris Diderot, Sorbonne Paris Cité, F-91191 Gif-sur-Yvette, France
    \and 
    Department of Space, Earth and Environment, Chalmers University of Technology, Onsala Space Observatory, 439 92 Onsala, Sweden
    \and
    Research Center for Space and Cosmic Evolution, Ehime University, 2-5 Bunkyo-cho, Matsuyama, Ehime 790-8577, Japan
    \and
    Department of Astronomy, University of Geneva, ch. d'Écogia 16, CH-1290 Versoix, Switzerland
    \and
    Academia Sinica Institute of Astronomy and Astrophysics, 11F of Astronomy-Mathematics Building, AS/NTU, No.1, Section 4, Roosevelt Road, Taipei 10617, Taiwan
    \and
    Università di Bologna, Dip. di Fisica e Astronomia “A. Righi”, Via P. Gobetti 93/2, I-40129 Bologna, Italy    
    }

\date{Received date /
Accepted date }
  \abstract
  {X-rays offer a reliable method to identify Active Galactic  Nuclei (AGN). However, in the high-redshift Universe X-ray AGN are poorly sampled due to their relatively low space density and the small areas covered by X-ray surveys. In addition to wide-area X-ray surveys, it is important to have deep optical data in order to locate the optical counterparts and determine their redshifts. In this work, we build a high-redshift ($z\geqslant3.5$) X-ray selected AGN sample in the XMM-XXL northern field using the most updated [0.5-2 keV] catalogue along with a plethora of new spectroscopic and multi-wavelength catalogues, including the deep optical Subaru Hyper Suprime-Cam (HSC) data reaching magnitude limits i$\sim$26 mag. We select all the spectroscopically confirmed AGN and complement this sample with high-redshift candidates that are HSC g- and r-band dropouts. To confirm the dropouts, we derive their photometric redshifts using spectral energy distribution techniques. We end up with a sample of 54 high-z sources (28 with spec-z), the largest in this field so far (almost three times larger than in previous studies), and we estimate the possible contamination and completeness. We calculate the number counts (logN-logS) in different redshift bins and compare our results with previous studies and models. We provide the strongest high-redshift AGN constraints yet at bright fluxes ($\rm f_{0.5-2~keV} > 10^{-15} erg~s^{-1}~cm^{-2}$). The samples of $z\geqslant3.5$, $z\geqslant4$ and $z\geqslant5$ are in agreement with an exponential decline  model similar to that witnessed at optical wavelengths. Our work emphasizes the importance of using wide area X-ray surveys with deep optical data to uncover high redshift AGN.}


   \maketitle
%

\section{Introduction}
The majority of massive galaxies in the local Universe host a super-massive black hole (SMBH) in their centre \citep{magorrian1998,kormendy2004,filippenko2003,barth2004,greene2004,greene2007,dong2007,greene2008}. The masses of these SMBHs vary between $10^5$ and $10^{10}$ solar masses. The accretion of matter onto the SMBH releases huge amounts of energy across the whole electromagnetic spectrum from the so-called Active Galactic Nucleus (AGN). Even though many studies suggest a correlation between the evolution of galaxies and SMBHs \citep[e.g.][]{silk1998,granato2004,dimatteo2005,croton2006,hopkins2006,hopkins2008,menci2008}, the physical processes lying behind such a correlation are not fully understood. Thus, a complete AGN sample including a diversity of physical properties (wide range of luminosity, mass, etc.) both at low and high redshifts is essential to understand better the SMBH evolutionary models and whether SMBHs play a role in the properties of their host galaxies. One of the most efficient ways to detect AGN is through X-ray emission, since X-rays penetrate the dust and gas surrounding the black hole without being absorbed, thus selecting both low-luminosity or/and moderately obscured black holes. This can be shown in the deepest X-ray fields, the Chandra Deep Field South (CDFS), where the number density of X-ray selected AGN is about 30,000 ${\rm deg^{-2}}$ with a median redshift value of $z=1.58\pm0.05$ \citep{luo2017}. In contrast, the number density of optically selected AGN is 100 times lower, $\sim300\, {\rm deg}^{-2}$ \citep{ross2013}.

However, this picture at high redshift  is reversed: the AGN samples are dominated by optically identified quasars compared to the poorly sampled X-ray selected sources. In particular, the number of known quasars within the first billion years of the Universe has been increasing rapidly over the last years, with tens of thousands optically selected AGN with $z\textgreater3$, and over 200 detected at $z\textgreater6$ \citep{banados2016,matsuoka2019,wolf2021}. At higher redshifts ($z\geqslant7$) only a small number of sources have been detected \citep{mortlock2011,banados2018,wang2018,matsuoka2019b} with the most distant objects found at $z=7.5$ \citep{banados2018,yang2020b} and $z=7.642$ \citep{wang2021}. Moreover, the optical surveys find that the normalization of the luminosity function of AGN presents an exponential drop at $z\geqslant3$ \citep{masters2012}. This could signpost the continuous creation of new AGN from a redshift of $z\sim20$ \citep{volonteri2010} to redshifts up to $z=2$, the `cosmic noon' era. Alternatively, the observed drop could be an artefact of reddening associated with the large amounts of gas that are abundant at these early cosmic epochs. On the other hand, just a few X-ray AGN have been found at these cosmic distances. This is because AGN are rare and large cosmic volumes or, equivalently, large sky areas are needed to find them. This is possible with the all-sky optical surveys, while X-ray surveys have covered significantly less area for equivalent depths. 

In the last years, efforts were made to compile high-z samples with dedicated X-ray surveys. \citet{vito2014} identified in total 141 X-ray sources in the redshift range $3\leqslant z \leqslant5.1$ in the 4 Ms Chandra Deep field (CDF) South \citep{xue2011}, the Chandra-COSMOS \citep{elvis2009} and the Subaru/XMM-Newton Deep Survey \citep[SXDS,][]{furusawa2008,ueda2008} fields. In the SXDS field (that overlaps with the XMM-XXL northern field with an area of 1.14\,deg$^2$), there were 30 high-z sources. 20/30 have spectroscopic redshifts from \citet{hiroi2012}. \citet{georgakakis2015} obtained about 340 X-ray sources in total with both spectroscopic and photometric redshifts in various fields observed either with Chandra or XMM-Newton X-ray telescopes. In the XMM-XXL northern field, they identified 55 (20) X-ray sources at $3\leqslant z \leqslant5$ ($z\geqslant3.5$) with only spectroscopic redshifts. In the Chandra Cosmos Legacy survey, \citet{marchesi2016} compiled a sample of 174 sources with 87 of them having available spectroscopic information ($3\leqslant z \leqslant5.3$).

More recently, \citet{vito2018} using the deep X-ray observations in the 7Ms CDF-South and 2 Ms CDF-North fields identified 102 high-z ($3\leqslant z \leqslant6$) AGN, while \citet{khorunzhev2019} selected a sample of 101 unabsorbed high-luminous quasars ($3\leqslant z \leqslant5.1$) using the 3XMM-DR4 catalogue \citep{watson2009}. At higher redshifts ($z\geqslant5$), in small and deep fields there are only three X-ray sources: two sources in the COSMOS field \citep{marchesi2016} with the highest one at $z=5.3$ \citep{capak2011}, and one source in the Chandra Deep Field North \citep{barger2003} at $z=5.186$. In the larger and shallower XMM-XXL field, \citet{menzel2016} found a source at $z=5.011$, while more recently, with the eROSITA telescope \citep{predehl2021} on board the Spectrum-Roentgen-Gamma mission, it was possible to identify one more source at $z=5.46$ \citep{khorunzhev2021}. In addition to the aforementioned studies, there are plenty of known high-z optically selected AGN matched with X-ray catalogues or follow-up X-ray observations \citep{vito2016,medvedev2020,wolf2021}. Even though they have not been selected through all-sky or dedicated X-ray surveys, their contribution is crucial to put some lower limits on the AGN space density.

In this work, we focus on selecting a sample of X-ray AGN in the early Universe in the XMM-XXL northern field \citep[][hereafter XXL Paper I]{pierreXXL} that has excellent multi-wavelength follow-up observations from the ultraviolet (UV) to the infrared (IR). To achieve this, we build a catalogue of spectroscopically confirmed AGN searching in the publicly available databases and we complement it with high-z sources selected through optical colour-colour criteria. We validate the colour-selected candidates through spectroscopic and photometric redshifts that we derive via spectral energy distribution (SED) fitting. Thanks to the Hyper Suprime-Cam \citep[HSC,][]{miyazaki2018} data that have deeper photometry, we aim at selecting a large sample of $z\geqslant3.5$ sources and obtaining a better constraint on the AGN sky-density distribution for the high-z population, at least for relatively high fluxes. Compared to the previous work of \citet{georgakakis2015} in the same field, we make use of the most up-to-date X-ray catalogue available that includes additionally the XMM-Newton observations occurred after 2012, increasing this way the total surveyed area by $\sim$40\%. 
Furthermore, besides the new spectroscopic data from the Sloan Digital Sky Survey (SDSS) IV and other surveys, we derive the photometric redshifts of all the colour-selected candidates increasing this way the number of the high-z sources. An additional advance is that the HSC data covering the area are much deeper than the SDSS or the Canada-France-Hawaii Telescope Legacy surveys, reaching magnitudes down to r$\simeq$26 mag \citep{aihara2019}. These depths are critical for the identification of AGN and, especially, the most obscured AGN where the nucleus is covered by veils of dust and gas and only the galaxy remains visible. To this end, we will be able to put much stronger constraints in the AGN sky-density and compare directly for the first time the theoretical population synthesis models at this redshift regime for bright fluxes. 

The data used in this work are presented in Sect.~\ref{data}. These include the X-ray catalogue and the multi-wavelength information along with the spectroscopic catalogues. In Sect.~\ref{sample}, we present the high-z sample selected through optical spectroscopy and the Lyman Break selection criteria. We also derive the photometric redshifts using SED fitting and we consider the contamination and reliability of our sample. In Sect.~\ref{counts}, we calculate the cumulative numbers in different redshift bins and constrain the bright flux of the logN-logS relation. In Sect.~\ref{summary}, we discuss and summarize the results. Throughout the paper, we assume a $\Lambda$CDM cosmology with H\textsubscript{0}=70 km s\textsuperscript{-1} Mpc\textsuperscript{-1}, $\Omega$\textsubscript{M}=0.3 and $\Omega$\textsubscript{$\Lambda$}=0.7.


\section{Data}\label{data}
   In this section, we describe the data used in this work to select high-z X-ray AGN in the XMM-XXL Northern field. We use the X-ray catalogue that was derived from the latest XMM-XXL pipeline and the deep HSC data in order to build the broad-band optical colours. Moreover, we use all the available spectroscopic data covering the field, while for the SED construction we use, in addition to HSC data, the ancillary data accompanying the X-ray catalogue that contains rich multi-wavelength data from UV to the mid-IR bands. The SEDs were used to estimate the photometric redshifts (photo-z) for those objects lacking spectroscopic redshift (spec-z). Below, we give a brief description of the aforementioned data and their surveys.

\subsection{XMM-XXL northern catalogue}\label{xrays}

The XMM-XXL survey (XXL Paper I) is the largest XMM-Newton programme approved (>6 Ms) surveying two extra-galactic sky regions of approximately equal size totaling $\sim~50\, {\rm deg^2}$ with a median exposure time of about 10\,ks per XMM-Newton pointing and a depth (at 3$\sigma$) of $\sim$$5\times 10^{-15}  {\rm erg~s^{-1}~cm^{-2}}$ in the 0.5-2 keV X-ray band. The X-ray data used in this study rely on an internal release obtained with the V4.2 XXL pipeline. The reader should be aware that the V4.2 version is expected to be superseded by the final catalogue, V4.3. Compared to the previous version V3 where the XMM observations were treated individually, the V4 version of the pipeline processes all the co-added observations together into $1\times1\,{\rm deg}^2$ mosaics, which enhances the detection sensitivity at any position. Furthermore, this version of the catalogue includes observations performed after 2012. These include in total 1.3 Ms observations with a median PN exposure time of $\sim$46 ks covering the XMM-Spitzer Extragalactic Representative Volume Survey (XMM-SERVS) field with an area of 5.3\,deg$^2$ \citep{chen2018}. In the analysis, we used the data from the equatorial sub-region of the field (4XMM-XXL-Northern; 4XXL-N) centred at R.A.$\sim$2h16m, Dec.$\sim$-4\degr52\arcmin  that covers an area of about 25\,deg$^2$ and contains 15547 X-ray sources \citep[][hereafter XXL Paper XXVII]{chiappettiXXL}. Restricting our sample to sources detected in the soft band, we ended up with 13742 X-ray sources. In the considered redshift range ($z\geqslant3.5$), the [0.5-2] keV energy band corresponds to rest-frame energies greater that [2.25-9] keV.

The 4XXL-N was accompanied with a multi-wavelength catalogue covering the spectrum from UV up to the mid-IR bands. In particular, it includes UV data from the GR6/7 release of the Galaxy Evolution Explorer survey \citep[GALEX,][]{bianchi2014} and optical data from the T0007 data release \citep{hudelot2012} of the Canada-France-Hawaii Telescope Legacy Survey (CFHTLS) and the 10$^{th}$ data release of the Sloan Digital Sky Survey \citep[SDSS,][]{ahn2014}. In the near-IR, the XXL-N field was covered by three European Southern Observatory (ESO) surveys with the Visible and Infrared Survey Telescope for Astronomy \citep[VISTA,][]{emerson2006}: the VISTA Hemisphere Survey \citep[VHS,][]{mcmahon2013}, the VISTA Kilo-degree Infrared Galaxy Survey \citep[VIKING,][]{edge2013} and the VISTA Deep Extragalactic Observations survey \citep[VIDEO,][]{jarvis2013}. Additionally, the multi-wavelength catalogue includes near-IR data from the UKIRT Infrared Deep Sky Survey \citep[UKIDSS,][]{dye2006} and the WIRcam camera on CFHT in the $K_s$ band \citep{moutard2016}. Finally, the catalogue was complemented with mid-IR photometry from the ALLWISE data release of the Wide-field Infrared Survey Explorer \citep[WISE,][]{wright2010} all-sky survey and the observation from the Infrared  Array  Camera \citep[IRAC,][]{fazio2004} on board the Spitzer Space Telescope \citep{werner2004}. The X-ray sources in the 4XXL-N were assigned a counterpart from this catalogue if there are detections in at least one band. The source matching was performed using the likelihood ratio estimator \citep{sutherland1992}. All of the X-ray sources were matched with existing catalogues. In particular, a counterpart was found for about 86\% of the sources in the optical, followed by 9.2\% and 4.4\% in near-IR and mid-IR or UV datasets, respectively. More details in the cross-matching techniques and the photometric data can be found in \citet[][hereafter XXL paper VI]{fotopoulouXXL} and XXL Paper XXVII.

\subsection{HSC-PDR2 catalogue}\label{wisedata}

We make use of the optical imaging data obtained with the Subaru Hyper-Suprime Camera (HSC) that is much deeper (i$\sim$26 mag) compared to the SDSS and CFHT optical surveys with magnitude limits (at 5$\sigma$) of i$\sim$21.3 mag and i$\sim$24.5 mag, respectively. More specifically, we use the second public data release \citep[HSC-PDR2,][]{aihara2019} of the Hyper Suprime-Cam Subaru Strategic Program \citep[HSC-SSP,][]{aihara2018b}. The full description of the HSC-SSP survey can be found in \citet{aihara2019}. Although the HSC-PDR2 data are taken in three layers with different area and depths (Wide, Deep, and UltraDeep), we only use the Wide field layer in this study. The overlapping area between 4XXL-N and HSC-PDR2 is approximately $\sim$24 sq. degrees, while the HSC-PDR2 5$\sigma$ sensitivity limits in the field reach mag 26.6, 26.2, 26.2, 25.3 and 24.5 in the AB magnitude system for the g, r, i, z and y bands, respectively.

In order to get a clean photometry, we followed the procedure described in the HSC-PDR2 website. Thus, we selected only the primary sources and we excluded those flagged with bad pixels, saturation or interpolation, hit by cosmic rays or located at the edges of the detector. Furthermore, we used the PDR2 masks centred around bright stars obtained by the GAIA DR2 catalogue. About 20\% of the sources were masked. To separate point-like and extended sources, we made use of the moments \citep{akiyama2018} in the i-band, because the i-band image has the highest image quality among the five bands in the HSC-SSP survey data \citep{aihara2019}. We use the PSF magnitudes for the point-like sources and the CMODEL magnitudes for the extended ones, which were estimated by fitting the PSF model and a two-component model \citep{abazajian2004,bosch2018}, respectively.

\subsection{4XXL-HSC sample}\label{4xxlHSC}

The common area between the HSC data (excluding the masked areas) and the 4XXL-N is about 20\,deg$^2$. This area includes 10998/13742 ($\sim$80\%) X-ray sources detected in the soft band. The HSC catalogue was cross-matched with the list of the X-ray sources using the coordinates of the ancillary data in a similar way to XXL paper VI (Sect. 4.1). In particular, we used a simple positional cross-matching method with a search radius depending on the data matched. Firstly, we cross-matched the HSC sources with the optical coordinates of the multi-wavelength catalogue with a radius of 1$\arcsec$. For those X-ray sources without an optical counterpart, we used the coordinates of the near-IR bands (1$\arcsec$) followed by the coordinates of the mid-IR and UV bands (2$\arcsec$). We ended up with 9689/10998 ($\sim$88\%) X-ray sources with HSC counterparts (hereafter 4XXL-HSC). Out of these, 7169 and 2520 sources are optically extended and point-like, respectively.

\begin{figure}
   \begin{tabular}{c}
    \includegraphics[width=0.45\textwidth]{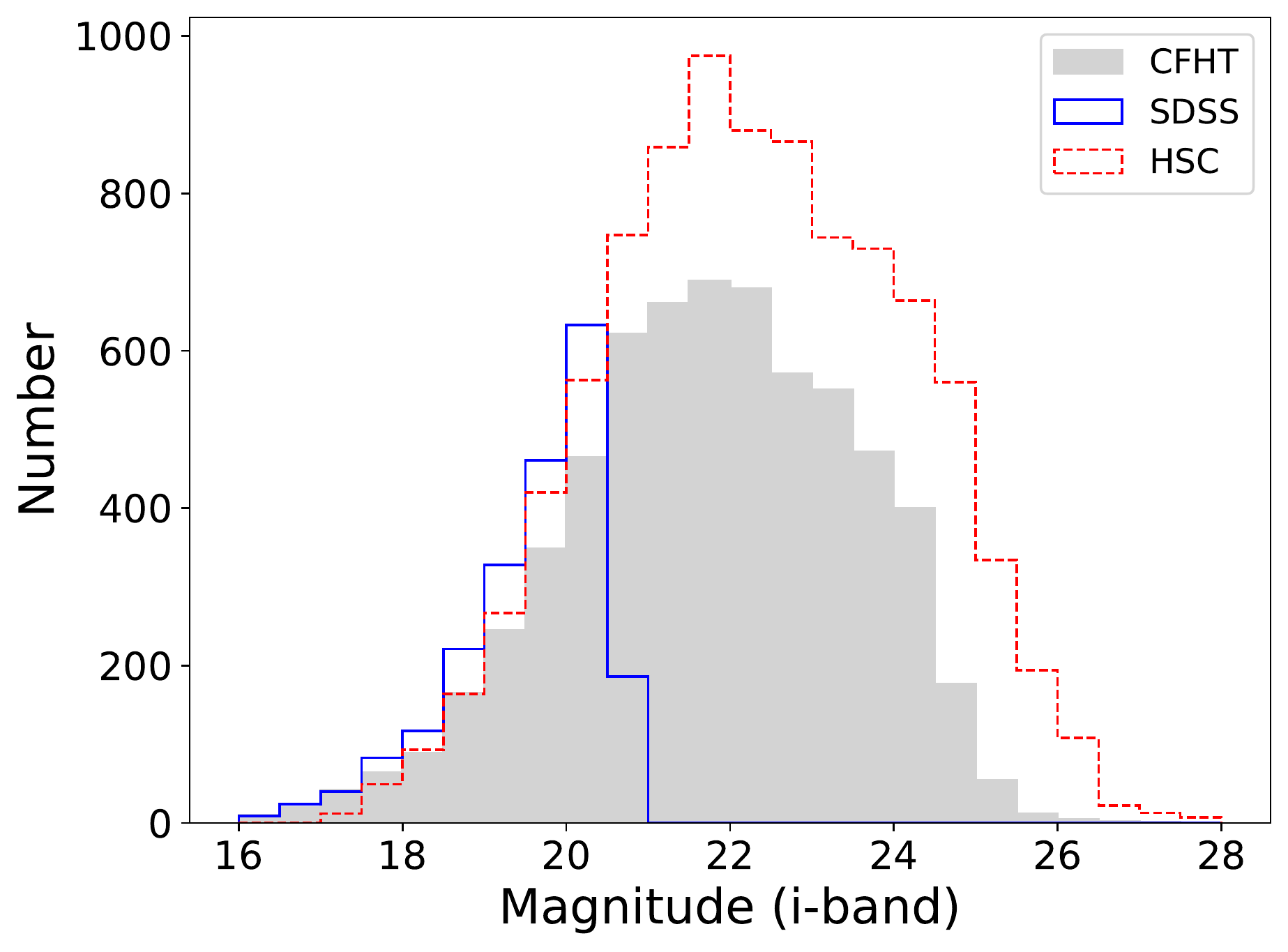}
    \end{tabular}
\caption{The magnitude distributions of the 4XXL-HSC sources for the CFHTLS (filled gray), SDSS (solid blue) and HSC (dashed red) i bands.} \label{magHist}
\end{figure}

In Fig.~\ref{magHist}, we show the 4XXL-HSC magnitude distributions in the i-band from the HSC data along with the SDSS and CFHT data for comparison. Including the HSC data, we go almost a magnitude deeper compared to the CFHT photometry and we have a higher number of optical counterparts. Moreover, the uncertainties of the HSC data are much lower for the fainter objects, thus our colour-selection criteria and photometric redshift estimations will be more precise.

\subsection{Spectroscopic catalogues}\label{zspeccatalogues}

In the XMM-XXL northern field, there are plenty of spectroscopic surveys targeting extra-galactic sources, both galaxies and AGN. Most of them target sources pre-selected in the UV or optical wavelengths, while there are some dedicated only to X-ray selected sources. In our analysis, we used the spectroscopic catalogues of X-ray selected AGN by \citet{menzel2016} and \citet{hiroi2012}. Moreover, we use the spectroscopic data gathered by the HSC team. In particular, they include the PRIsm MUlti-object Survey \citep[PRIMUS][]{coil2011,cool2013} in the sub-region XMM-LSS ($\sim$2.88\,deg$^2$) of the XXL-N, the Galaxy And Mass Assembly \citep[GAMA,][]{liske2015}, the VIMOS VLT Deep Survey \citep[VVDS,][]{lefevre2013} and the VIMOS Public Extragalactic Survey \citep[VIPERS,][]{garilli2014}. Also, in this database the DR12 \citep{alam2015} and DR14 \citep{paris2018} of the SDSS are included. Additionally, we used the latest data release \citep[SDSS-DR16,][]{ahumada2020} that is the fourth release of the Sloan Digital Sky Survey IV. The spectroscopic information provided by the HSC team was already associated with the photometric catalogue of the sources. For the remaining datasets, the spectroscopic catalogues were matched to the optical positions in our sample with a radius of 1$\arcsec$. For all the sources, we selected high quality flags that correspond to probability of 90\% or higher that redshift is the true one.


\section{Sample selection}\label{sample}

In this section, we present the final sample of the high-z sources in the XXL-N field. We list the confirmed high-z sources found in the publicly available spectroscopic catalogues and also the colour-selected AGN. To account for any contamination by low-z interlopers, we use the spectroscopic redshifts and for the remaining sources we estimate the photo-z with the \texttt{X-CIGALE} algorithm \citep{yang2020}. Furthermore, we use the X-ray to optical flux ratio as an additional criterion to account for brown dwarf contamination. 


\subsection{Spectroscopic redshifts}\label{specz}

Using the spectroscopic catalogues mentioned in Sect.~\ref{zspeccatalogues}, we select initially all the X-ray sources with $z\geqslant3.5$. In particular, from the latest data release of SDSS (DR16) \citep{ahumada2020} we select 37 objects. Also, we make use of the spectroscopic information provided by the HSC PDR2 database and select 27 sources at high redshift and assigned with a secure flag. Furthermore, we include 19 and 8 sources with spectroscopic redshifts found in \citet{menzel2016} and \citet{hiroi2012}, respectively. It is worth mentioning that there are two sources with spectroscopic redshift in \citet{menzel2016} without X-ray counterparts in the 4XXL catalogue. This is probably due to different source detection algorithms or background estimations used. The first one lies at $z=3.67$, while the second one is the most distant X-ray source in their catalogue at $z=5.011$. Both sources are broad-line AGN of type 1 according to their optical spectra \citep{liu2016} and do not show strong absorption in the X-ray regime ($\rm logN_H<21.3$). These sources are not taken into consideration in this study. The discrepancies between the two X-ray catalogues are out of the scope of this study and will not be discussed further.

In total, taking into account the overlaps within these catalogues, we end up with 45 sources at high redshifts. Even though all these sources have secure measurements according to the flags provided, we visually inspected all the individual spectra for probable outliers or having low quality. We excluded nine sources, since their spectra are too noisy, resulting in 36 spec-z sources. Out of those, 28 spec-z sources are falling in good regions (outside of the HSC masked areas). 22/28 (79\%) are point-like sources, while six sources appear extended in the HSC optical images.


\subsection{High-z dropout candidates}\label{4.1}

\begin{figure}
   \begin{tabular}{c}
    \includegraphics[width=0.49\textwidth]{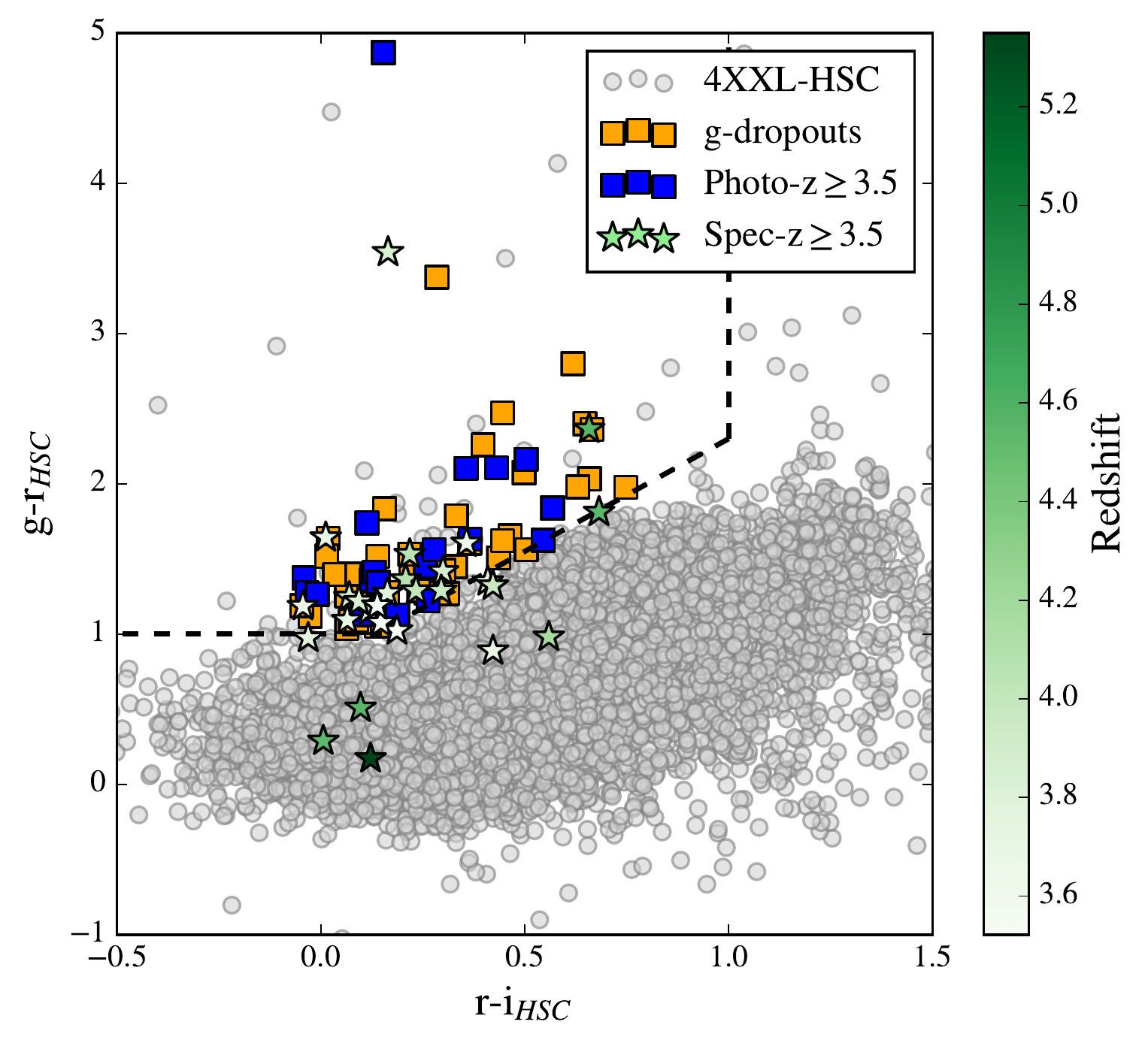} \\
    \includegraphics[width=0.49\textwidth]{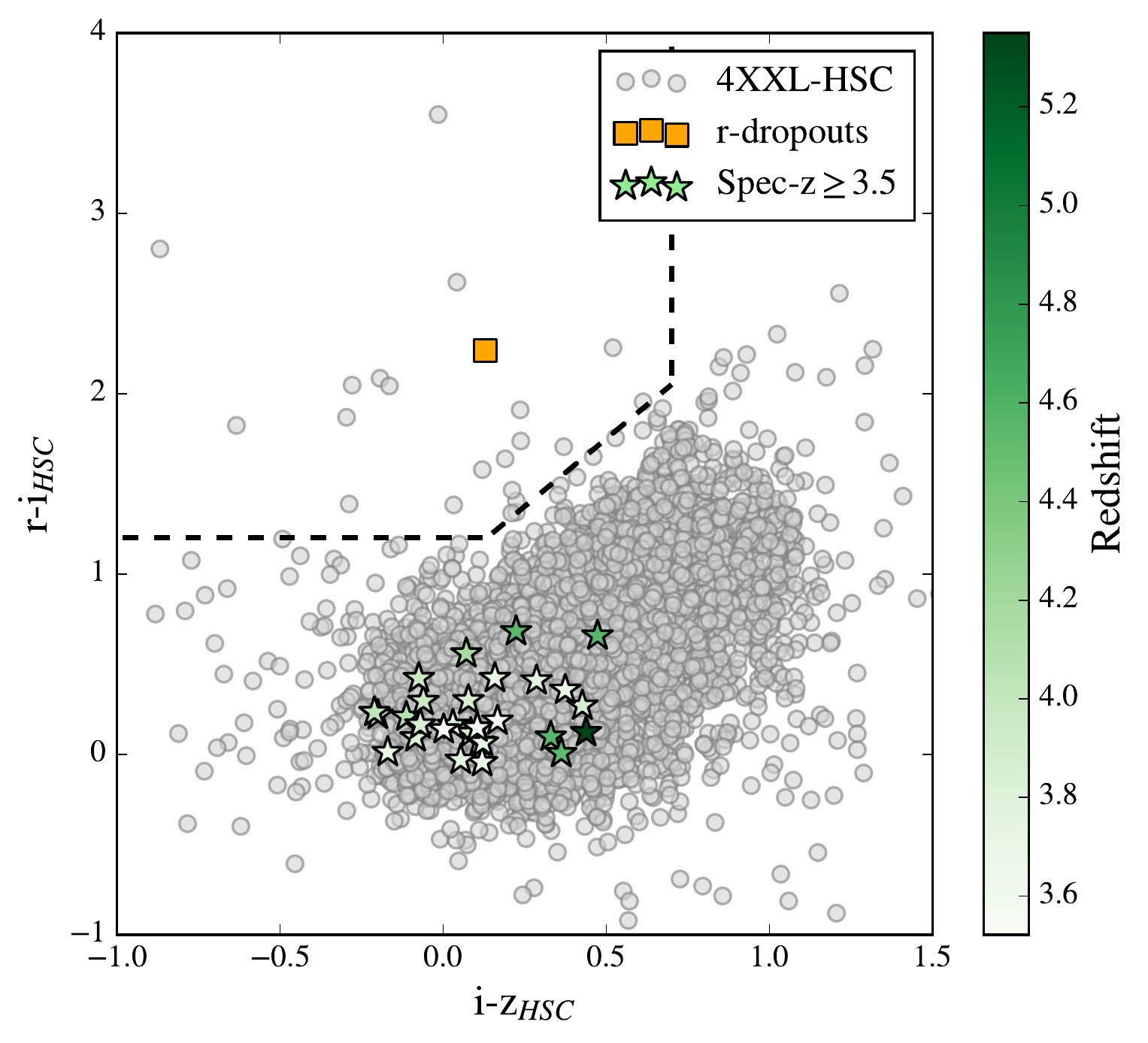}
    \end{tabular}
\caption{The (g-r, r-i) and (r-i, i-z) colour-colour plots (from top to bottom). The black lines indicate the selection criteria defined by \citet{ono2018}. The small gray points indicate the full 4XXL-HSC sample and the orange squares represent the high-z candidates. We over-plot the specz-z sample (asterisks) colour-coded with the redshift, while we highlight in blue colour the dropouts with $z_{phot}\geqslant3.5$.} \label{onoPlots}
\end{figure}

We complement the spectroscopically confirmed sources by using the Lyman Break colour selection method for the 4XXL-HSC sample. The Lyman Break technique \citep{steidel1996,steidel1999,giavalisco2002} has been widely used to select high-z sources using UV, optical or IR filters. In this work, we use a set of colours and criteria similar to \citet{ono2018} for both point-like and extended sources in order to select candidates at $z\sim4-7$. Sources at $z=$4, 5, 6 and 7 are expected to be selected by the gri, riz, izy and zy colours, respectively. In brief, we applied the following criteria to the 4XXL-HSC sample:

For g-dropouts (z$\sim$4):  
\begin{ceqn}
\begin{equation}g-r>1.0,\, r-i<1.0
\end{equation}
\end{ceqn}
\begin{ceqn}
\begin{equation}
g-r>1.5\times(r-i)+0.8 
\end{equation}
\end{ceqn}
\begin{ceqn}
\begin{equation}(S/N)_i>5
\end{equation}
\end{ceqn}

For r-dropouts (z$\sim$5):  
\begin{ceqn}
\begin{equation}
r-i>1.2,\, i-z<0.7
\end{equation}
\end{ceqn}
\begin{ceqn}
\begin{equation}
r-i>1.5\times(i-z)+1.0
\end{equation}
\end{ceqn}
\begin{ceqn}
\begin{equation}
(S/N)_z>5 \:{\rm and} \:(S/N)_g<2
\end{equation}
\end{ceqn}

For i-dropouts (z$\sim$6): 
\begin{ceqn}
\begin{equation}
i-z>1.5,\, z-y<0.5,
\end{equation}
\end{ceqn}

\begin{ceqn}
\begin{equation}
i-z>2.0\times(z-y)+1.1
\end{equation}
\end{ceqn}

\begin{ceqn}
\begin{equation}
(S/N)_z>5 \:{\rm and}\: (S/N)_y>4
\end{equation}
\end{ceqn}

\begin{ceqn}
\begin{equation}
(S/N)_g<2 \:{\rm and}\: (S/N)_r<2
\end{equation}
\end{ceqn}

For z-dropouts (z$\sim$7):  
\begin{ceqn}
\begin{equation}
z-y>1.6 \:{\rm and} \:(S/N)_y>5.
\end{equation}
\end{ceqn}

These colours returned in total 69 high-z candidates, 68 g-dropouts and 1 r-dropout. We did not have any i or z dropouts. Out of these, there are 27 point-like and 42 extended sources. Figure~\ref{onoPlots} shows the colour-colour plots for g- and r-band dropouts. The orange points inside the wedges (dashed lines) represent the dropout sources in each selection method, while the gray points represent the whole 4XXL-HSC sources. There are some sources inside the wedges not selected as high-z candidates, since they did not meet the signal-to-noise detection threshold criteria mentioned above. 

\begin{figure}
   \begin{tabular}{c}
        \includegraphics[width=0.49\textwidth]{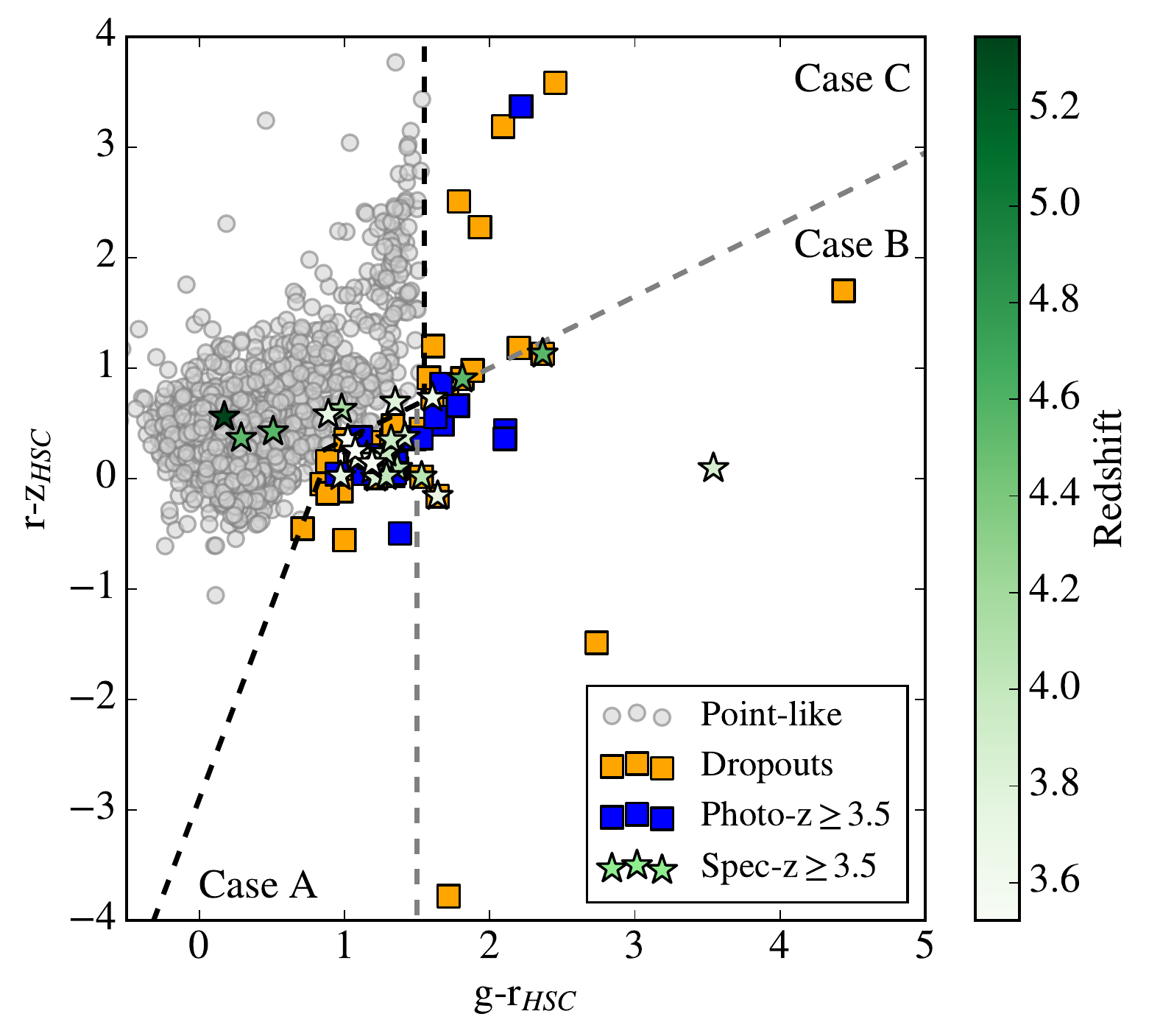}    \end{tabular}
\caption{The (g-r, r-z) colour-colour plot. The black lines indicate the selection criteria defined by \citet{akiyama2018} including the complementary criteria from this work. The small gray points indicate the point-like sources in the 4XXL-HSC sample, while the orange squares represent the high-z candidates in Cases A, B and C (see text for details). We over-plot the specz-z sample (asterisks) colour-coded with the redshift}, while we highlight in blue colour the dropouts with $z_{phot}\geqslant3.5$. \label{akiyamaPlot}
\end{figure}

Finally, we constructed, additionally to the above criteria, a high-z sample on the basis of the broad band colours described in \citet{akiyama2018}. The latter criteria that were optimised only for the point-like sources and especially for the high-z quasars rely on the different colours used (g-r vs. r-z) compared to the classic g-r vs. r-i. Furthermore, since \citet{akiyama2018} were interested in the redshift range of z$\sim$3.5-4.5, we relaxed their criteria including this way sources with higher redshifts. Thus, besides their original criteria (case A), we have included the area with g-r colour greater than 1.5 not to limit only sources up to z=4.5 (case B) and also a second area where there are quasars with much higher redshifts (case C). These additional criteria were based on the spectroscopic information of known quasars and stars as shown in \citet[][see their Fig. 2]{akiyama2018}. Below, we summarize these optimized criteria:

Case A:  
\begin{ceqn}
\begin{equation}
0.65\times(g-r)-0.30>(r-z)
\end{equation}
\end{ceqn}
\begin{ceqn}
\begin{equation}
3.50\times(g-r)-2.90>(r-z)
\end{equation}
\end{ceqn}
\begin{ceqn}
\begin{equation}
(g-r)<1.5
\end{equation}
\end{ceqn}
Case B:  
\begin{ceqn}
\begin{equation}
0.65\times(g-r)-0.30>(r-z)
\end{equation}
\end{ceqn}
\begin{ceqn}
\begin{equation}
3.50\times(g-r)-2.90>(r-z)
\end{equation}
\end{ceqn}
\begin{ceqn}
\begin{equation}
(g-r)\geq1.5
\end{equation}
\end{ceqn}
Case C:
\begin{ceqn}
\begin{equation}
3.50\times(g-r)-2.90>(r-z)
\end{equation}
\end{ceqn}
\begin{ceqn}
\begin{equation}
(g-r)\geq1.55
\end{equation}
\end{ceqn}

In Fig.~\ref{akiyamaPlot}, we plot the selected dropout sources in the g-r vs. r-z colours space. By applying these criteria, we were able to identify 33, 14 and 12 high-z candidates in cases A, B and C, respectively, resulting in 59 sources in total. Concerning the overlap between sources derived by the \citet{ono2018} and \citet{akiyama2018} methods, the latter recovered all the point-like sources, but selected additionally 32 sources. Thus, the addition of the extra criterion was critical to select a more complete high-z sample. In total, using both selection criteria of the aforementioned studies, we were able to build a sample of 101 unique high-z candidates.

The sources derived with the broad band selection methods are expected to be contaminated by a population of low redshift interlopers and also by brown dwarfs. Concerning the low-z interlopers, we use the spectroscopic information of the sources if available and for the remaining we derive the photometric redshift in the next section. Also, in our case the contamination from the stellar objects is expected to be negligible, since we are using X-ray selected sources that are well fitted with AGN templates (Sect.~\ref{sed}). In Sect.~\ref{contamination}, we address both these issues in detail.


\subsection{Photometric redshifts}\label{photoz}

Out of the final 101 dropouts, 31 have a counterpart in one of the spectroscopic catalogues. Out of these, 21 (68\%) sources have z$\mathrm{_{spec}}\geqslant3.5$, while 10 (32\%) sources have lower redshifts. The vast majority of the latter are located outside the \citet{ono2018} wedges in all colour diagrams, while they are all marginal to those of \citet{akiyama2018}. For the remaining 70 dropouts lacking spectroscopic information we derive the photometric redshifts using all the available data from UV to mid-IR wavelengths. In this section, we present the method used to derive the photometric redshifts and also the different statistical approaches to calculate its accuracy and reliability.


\subsubsection{SED fitting}\label{sed}
We perform a multi-component SED fitting with the \texttt{X-CIGALE} algorithm to estimate the photometric redshifts. \texttt{X-CIGALE} is the latest version of the Code Investigating GALaxy Emission \citep[\texttt{CIGALE},][]{noll2009,Ciesla2015,boquien2019} and has been used recently for redshift estimations in the early Universe with high precision \citep{barrufet2020,toba2020,shi2021}. For example, \citet{shi2021} found a catastrophic failure ratio of 10\% with normalized uncertainty $\sigma_{\rm NMAD}$=0.08. The \texttt{X-CIGALE} code fits the observational multi-wavelength data with a grid of theoretical models and returns the best-fitted values for the physical parameters. The results are based on the energy balance, i.e., the energy absorbed by dust in UV/optical is re-emitted after heating at longer wavelengths, such as the mid-IR and far-IR. 

For the SED fitting, we used all the available data covering the wavelength range from the ultraviolet light up to mid-IR bands. In our analysis, we used redshift values between 0.0 and 7.0 with a step of $\Delta z=0.05$, while we built a grid of models including different stellar populations, dust attenuation properties, dust emission, star formation history and AGN emission. With this configuration, for each source we fit the observational data to more than ~350 millions models. The models and the parameter space covered by the SED components are described below:

\begin{enumerate}
\item For the stellar population, we used the synthesis models of \citet{bruzual2003} with the Initial Mass Function (IMF) by \citet{salpeter1955} and a constant solar metallicity at Z\,=\,0.02. A constant metallicity does not affect significantly the shape of the SEDs \citep[and references therein]{pouliasis2020}.

\item For the attenuation originated in the absorption and scatter of the stellar and nebular emission by interstellar dust, we use the \citet{calzetti2000} attenuation law.

\item The emission by dust in the IR regime was modelled by the \citet{draine2014} templates, an updated version of the \citet{draine2007} models that allow to include higher dust temperatures. The main parameters that describes them are the mass of the PAH population, $q_{PAH}$, and the dust temperature which is expressed with the minimum radiation field, $U_{min}$ \citep{aniano2012}.

\item The AGN templates used in our analysis are based on the realistic clumpy torus model presented in \citep[SKIRTOR]{stalevski2012,stalevski2016}. SKIRTOR assumes a clumpy two-phase torus model that considers an anisotropic, but constant, disk emission. More details about the SKIRTOR implementation in X-CIGALE can be found in \citet{yang2020}. The parameter space for this module followed the description in \citet{yang2020} and \citet{mountrichas2021}. Moreover, with \texttt{X-CIGALE} it is possible to include the polar dust extinction \citep{mountrichas2021,toba2021}.

\item Finally, for the star formation history (SFH) we use a double-exponentially-decreasing model (2$\tau$-dec) model with different e-folding times.
\end{enumerate}

\begin{figure}
   \begin{tabular}{c}
    \includegraphics[width=0.46\textwidth]{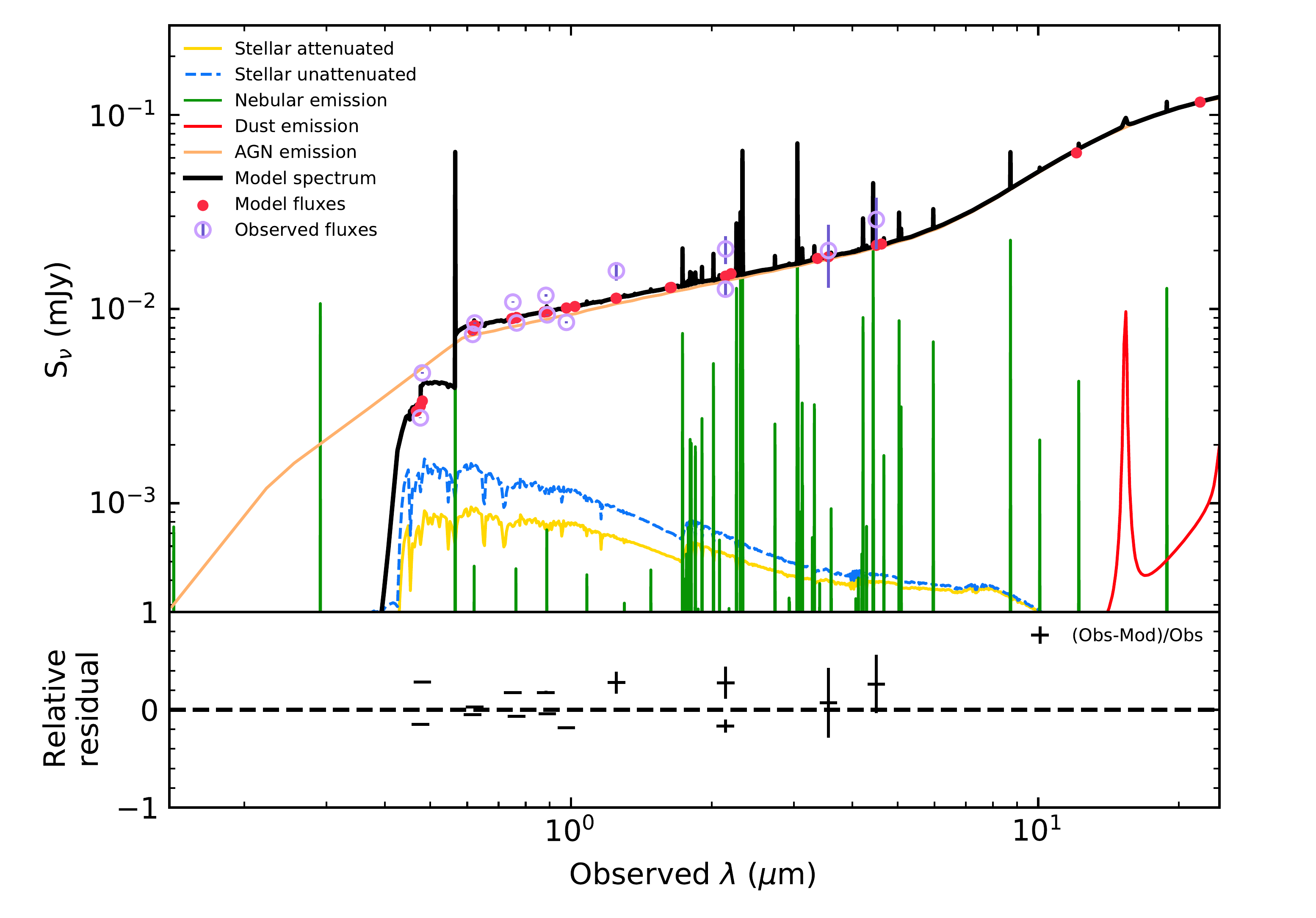} \\
    \includegraphics[width=0.43\textwidth]{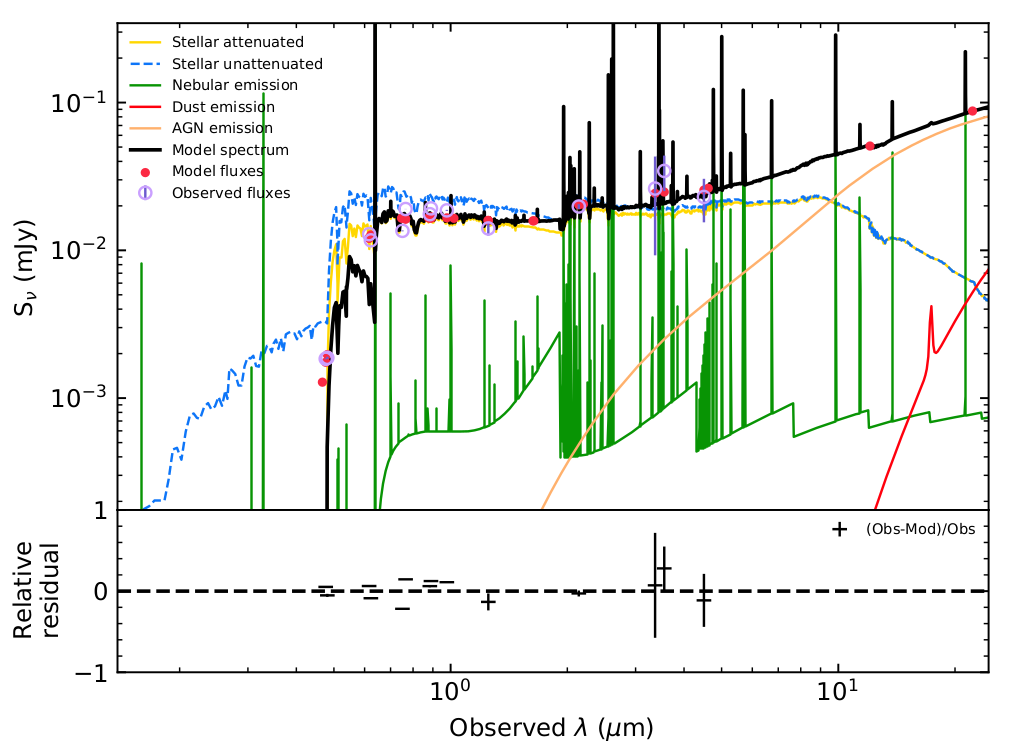} \\
    \includegraphics[width=0.43\textwidth]{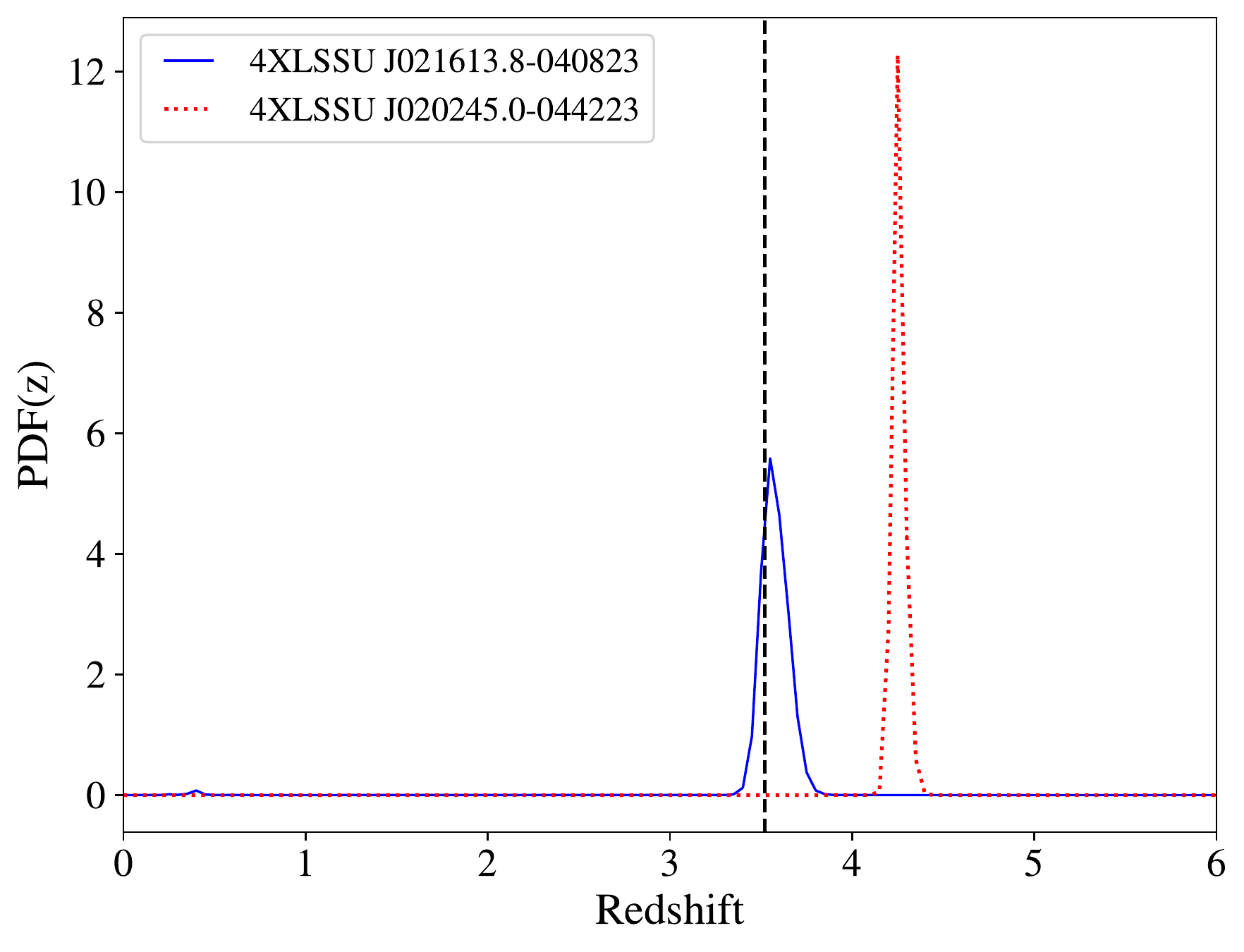}
    \end{tabular}
\caption{Examples of SED fits of a spec-z source (ID=J021613.8-040823, upper) and a photo-z source (ID=J020245.0-044223, middle). The dust emission is plotted in red, the AGN component in orange, the attenuated (unattenuated) stellar component is shown by the yellow (blue) solid (dashed) line, while the green lines shows the nebular emission. The total flux is represented with black colour. Below the SEDs, we plot the relative residual fluxes versus the wavelength. Lower panel: The probability density function of redshift, PDF(z), of the sources with a peak at z$\mathrm{_{peak}}=3.55$ and z$\mathrm{_{peak}}=4.25$, respectively. The vertical line shows the spectroscopic redshift of J021613.8-040823 at z$\mathrm{_{spec}}=3.522$.}\label{exampleSED}
\end{figure}

Table \ref{proposal} lists the models and the parameters space used in the SED fitting procedure. We removed seven sources (1 have spectroscopic redshift) with very high reduced $\chi^2$ values, $\chi^2_{red}>10$ \citep{mountrichas2021}, from the SED fitting process. These were mainly very faint sources detected in only a few bands and/or located in high source density areas. Finally, for each source, we obtained the full probability density function of the redshift, PDF(z). In Fig.~\ref{exampleSED} we give an example of the SED and the PDF(z) of the source with ID= J021613.8-040823 and z$\mathrm{_{spec}}=3.522$. The peak of the PDF(z) for this source is z$\mathrm{_{peak}}=3.55$.

\begin{table*}
\caption{The models and their parameter space used by \texttt{X-CIGALE} for the SED fitting of the colour-selected high-z sources.}
\begin{tabular}{ l c r }
\hline
\multicolumn{1}{l}{Parameter} &  & Value \\ \hline \hline
\multicolumn{3}{c}{Star formation history: double-exponentially-decreasing (2$\tau$-dec) model}\\
Age of the main stellar population in Myr && 500,1000,3000,5000,7000,9000 \\
e-folding time of the main stellar population model in Myr, $\tau_{\rm main}$ && 100,500,1000,3000,5000,9000 \\
Age of the late burst in Gyr, $age_{\rm burst}$ && 10,50,100,200,400  \\
$f_{\rm burst}$ && 0.0, 0.1, 0.3, 0.5\\
$\tau_{\rm burst}$ && 3000, 9000\\
\hline
\multicolumn{3}{c}{Stellar population synthesis model}\\
Single Stellar Population Library&&\citet{bruzual2003}\\
Initial Mass Function&& \citet{salpeter1955} \\
Metallicity && 0.02 (Solar) \\
\hline
\multicolumn{3}{c}{Nebular emission}\\
Ionization parameter ($\log U$)&& -2.0 \\
Fraction of Lyman continuum escaping the galaxy ($f_{\rm esc}$)&& 0.0 \\
Fraction of Lyman continuum absorbed by dust ($f_{\rm dust}$)&& 0.0 \\
Line width in km/s&& 300.0 \\
\hline
\multicolumn{3}{c}{Dust attenuation: \citet{calzetti2000}} \\
Colour excess of stellar continuum light for young stars E(B-V) && 0.05, 0.3, 0.5, 0.7, 0.9 \\
Reduction factor for the E(B-V) of the old stars compared to the young ones && 0.44\\
\hline
\multicolumn{3}{c}{Dust template: \citet{draine2014}}\\
Mass fraction of PAH ($\rm q_{\rm pah}$) && 0.47, 4.58, 7.32\\
Minimum radiation field ($\rm U_{min}$) && 5, 50\\
Powerlaw slope dU/dM propto $\rm U^\alpha$ && 2.0\\
Fraction illuminated from $\rm U_{\rm min}$ to $\rm U_{\rm max}$, $\gamma$ && 0.1\\
\hline
\multicolumn{3}{c}{AGN models from \citet{stalevski2016} (SKIRTOR)}\\
 Average edge-on optical depth at 9.7 micron (t) &&   7.0\\
 Power-law exponent that sets radial gradient of dust density (pl) && 1.0\\
 Index that sets dust density gradient with polar angle (q) && 1.0 \\
 Angle measured between the equatorial plan and edge of the torus (oa) && 40\\
 Ratio of outer to inner radius, $\rm R_{\rm out}/R_{\rm in}$ && 20\\
 Fraction of total dust mass inside clumps ($\rm M_{\rm cl}$) && 97\%\\
 Inclination angle ($i$) && 30, 70\\
 AGN fraction && 0.01,0.1,0.2,0.3,0.4,0.5,0.6,0.7,0.8,0.9,0.99 \\
 Extinction in polar direction, E(B-V) && 0.0, 0.8\\
 Emissivity of the polar dust && 1.6 \\
 Temperature of the polar dust (K) && 100.0\\
 The extinction law of polar dust && SMC\\
\hline
\multicolumn{3}{c}{}\\
 Redshift values && 0.0-7.0 with a step of 0.05  \\
\hline
\end{tabular}
\tablefoot{Edge-on, type-2 AGN have inclination $i=70$~degrees and face-on, type-1 AGN have $i=30^\circ$. The extinction in polar direction, E(B-V), included in the AGN module, accounts for the possible extinction in type-1 AGN, due to polar dust. The AGN fraction is measured as the AGN emission relative to IR luminosity (1--1000 $\mu$m).}
\label{proposal}  
\end{table*}

\subsubsection{Photo-z performance}\label{accuracy}

\begin{figure}
   \begin{tabular}{c}
    \includegraphics[width=0.45\textwidth]{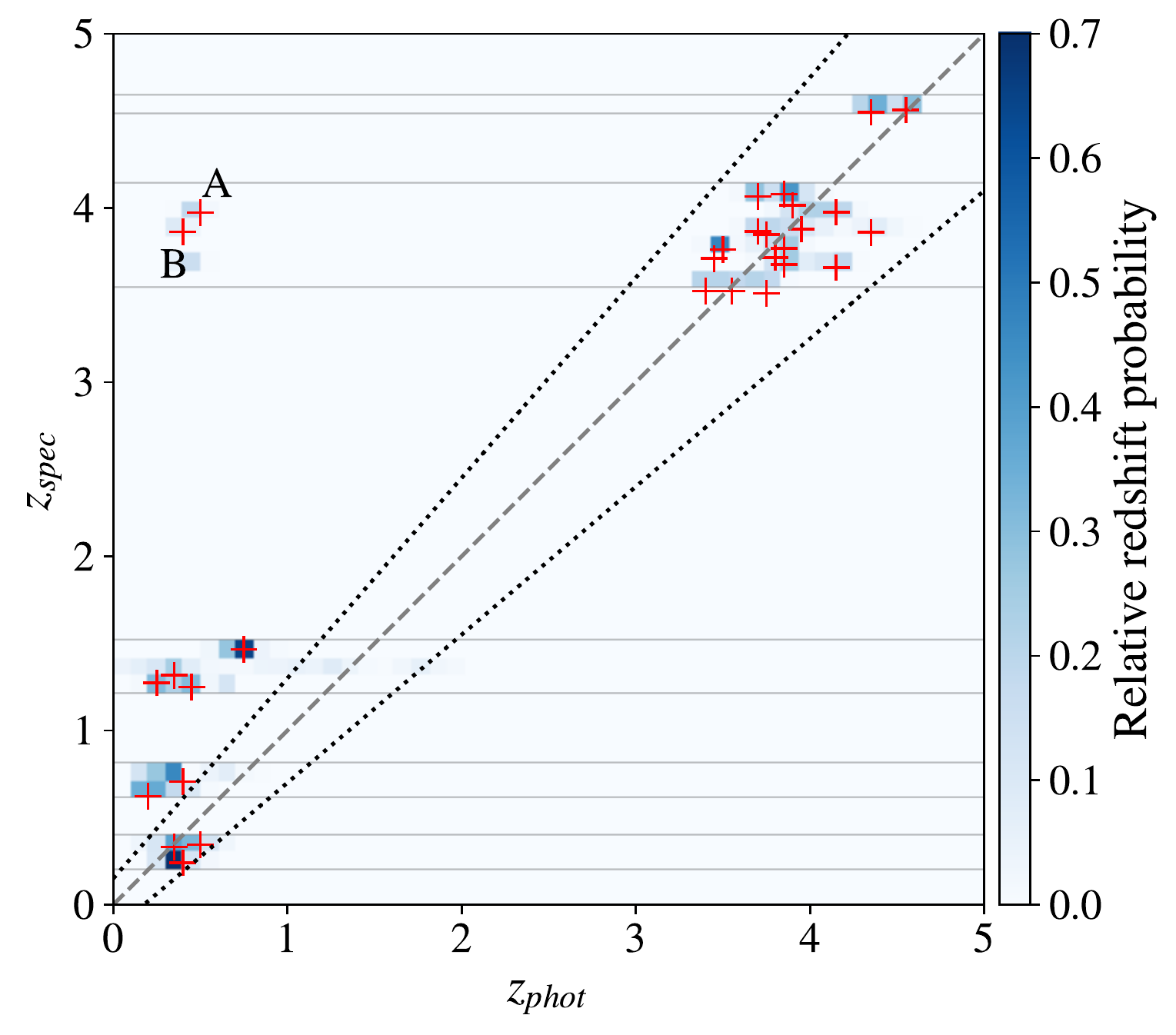} 
    \end{tabular}
\caption{Photometric versus spectroscopic redshifts for the 30 dropouts that have available spec-z information. The dotted lines represents the limits of the catastrophic outliers. The points are colour-coded with the relative redshift probability. Red crosses represent the peaks of the PDFs. A and B indicate the two extremes outliers, ID:  J021727.6-051718 and  J021952.8-055958, respectively.} \label{RedshiftCompare2}
\end{figure}

\begin{figure}
   \begin{tabular}{c}
   \includegraphics[width=0.47\textwidth]{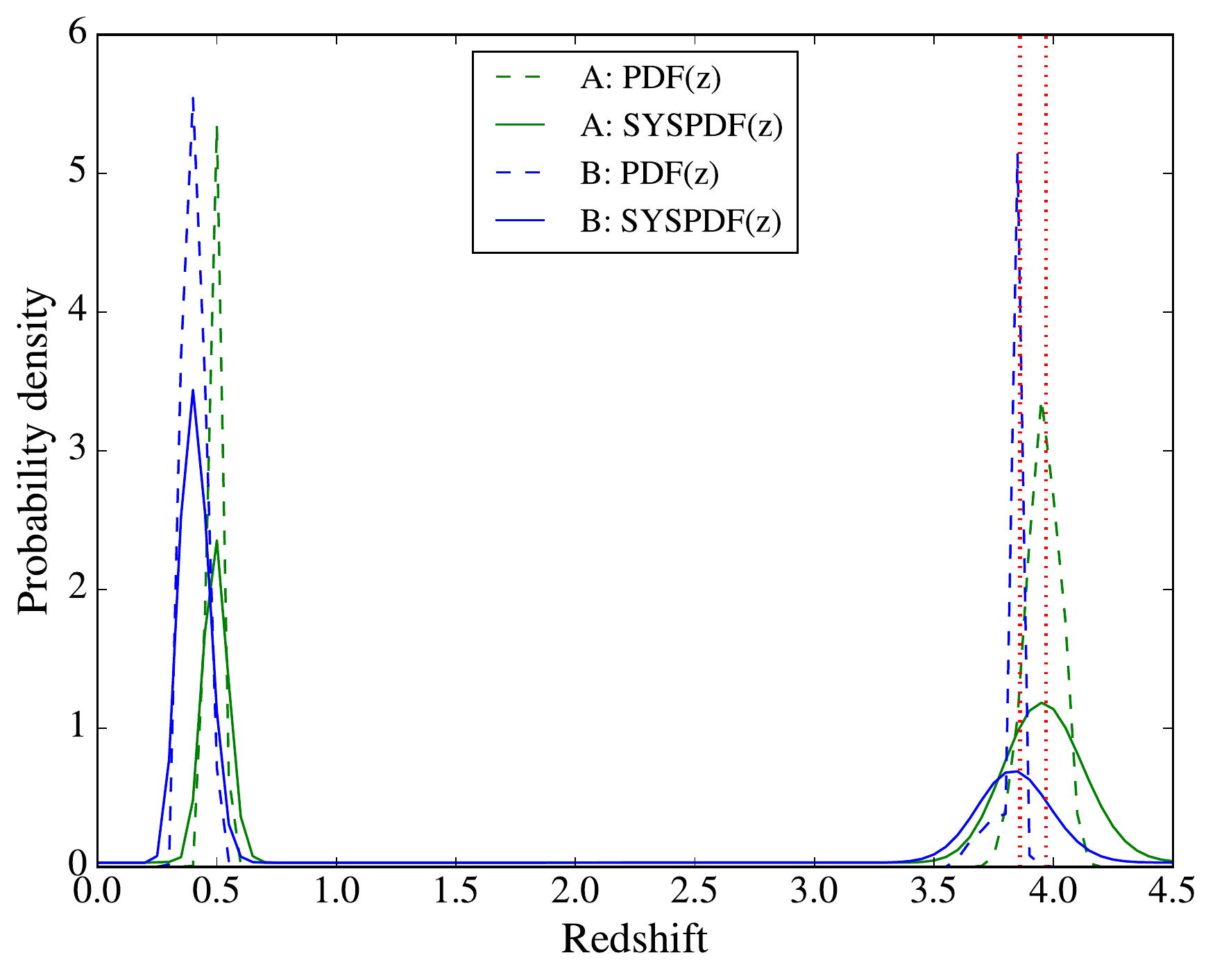} \\
    \end{tabular}
\caption{The PDF(z) of two sources,  J021952.8-055958 (blue dashed) and  J021727.6-051718 (green dashed), with spec-z 3.86 and 3.97 (vertical dotted lines), respectively. The corrected PDF(z) for systematics, SYSPDF(z), are shown with solid lines.} \label{outliers}
\end{figure}

The photometric redshift accuracy for the high-z candidates was investigated using the dropout sample of 30 sources with available spectroscopic information and low $\chi^2_{red}$ values. The scatter between the photometric and spectroscopic redshifts usually is estimated using the traditional statistical indicators: the normalised median absolute deviation $\sigma_{\rm NMAD}$ \citep{hoaglin1983, salvato2009,ruiz2018} and the percentage of the catastrophic outliers $\eta$ \citep{ilbert2006,laigle2016} defined as follows:

\begin{ceqn}
\begin{equation}
\sigma_{\rm NMAD}=1.4826\times median \left( \frac{|\Delta z-median(\Delta z)|}{1+z_{\rm spec}}\right)
\end{equation}
\end{ceqn}
\begin{ceqn}
\begin{equation}
\eta=\frac{N_{\rm outliers}}{N_{\rm total}}\times100
\end{equation}
\end{ceqn}
where $\Delta z=z_{\rm phot}-z_{\rm spec}$, $N_{\rm total}$ is the total number of sources and $N_{\rm outliers}$ is the number of the outliers. To be consistent with previous works in the literature, we define an object as an outlier if it has $|\Delta z|/(1+z_{\rm spec})>0.15$. We obtained the values $\eta$=26.7\% and $\sigma_{\rm NMAD}$=0.08 for the whole sample, but when considering only spec-z sources at high redshift ($z\geqslant3.5$), the situation significantly improves with $\eta$=9.5\% and $\sigma_{\rm NMAD}$=0.05. In Fig.~\ref{RedshiftCompare2}, we show the photo-z peak values (red crosses) as a function of the spec-z. There are two extreme outliers at high-z regime, ID: J021952.8-055958 and  J021727.6-051718, with spec-z equal to 3.86 and 3.97, respectively. We show the PDF(z) of these sources in Fig.~\ref{outliers} with dashed lines. Even though, they are spectroscopically confirmed high-z sources, the nominal value of the redshift is at z$\simeq$0.5, while there is a secondary peak around the true value. 

A caveat of the classic scatter estimation method is that it considers only the PDF most likely values (i.e. the mode) and not the PDF(z) as a whole. In order to estimate properly the systematic biases of the scatter and the percentage of outliers using all the information contained in the PDF(z) obtained with \texttt{X-CIGALE}, we followed a more updated and refined interpretation of the biases proposed by \citet[Appendix B,][]{buchner2015}. The idea is to find the $\tilde{\eta}$ and $\tilde{\sigma}$ values that maximize the likelihood that the true redshift of the source (spec-z) is given by the full PDF(z). The likelihood is given by a modified version of the PDF(z), with a term that is the broadening of the PDF(z) due to the scatter (convolution with a Gaussian with zero mean and  standard deviation, $\tilde{\sigma}$), plus a constant term due to the probability of being a catastrophic outlier: 

\begin{ceqn}
\begin{equation}\label{eqsys}
SYSPDF(z) :=  \tilde{\eta} \times U(0,z_{max})+(1-\tilde{\eta})\times (PDF*N(0,\tilde{\sigma}))
\end{equation}
\end{ceqn}

Then, the total likelihood to maximise is the product of the SYSPDF(z) value for the corresponding spec-z of each source:

\begin{ceqn}
\begin{equation}
L(\tilde{\sigma},\tilde{\eta})=\prod\limits_{i} SYSPDF_i(z_{spec},i).
\end{equation}
\end{ceqn}

Using a simple numerical minimization method for the likelihood, we find $\tilde{\eta}$=12.6\% and $\tilde{\sigma}$=0.04. However, for a better estimation of the parameters, we used the \texttt{UltraNest} code \citep{buchner2021}, a Bayesian posterior sampling method based on nested sampling \citep{skilling2004,skilling2009}. We assumed for both $\tilde{\eta}$ and $\tilde{\sigma}$ Jeffreys priors (i.e. a uniform prior in the logarithmic space) with limits (1e-4,1.0) and (1e-3,1.0), respectively. This way, we may have a complete characterization with the posterior probability distribution of the parameters. To this end, the systematics of our sample are given by $\tilde{\eta}=19.3^{+23.2}_{-14.4}\%$ and $\tilde{\sigma}=0.04^{+0.05}_{-0.02}$.

\begin{figure}
   \begin{tabular}{c}
    \includegraphics[width=0.47\textwidth]{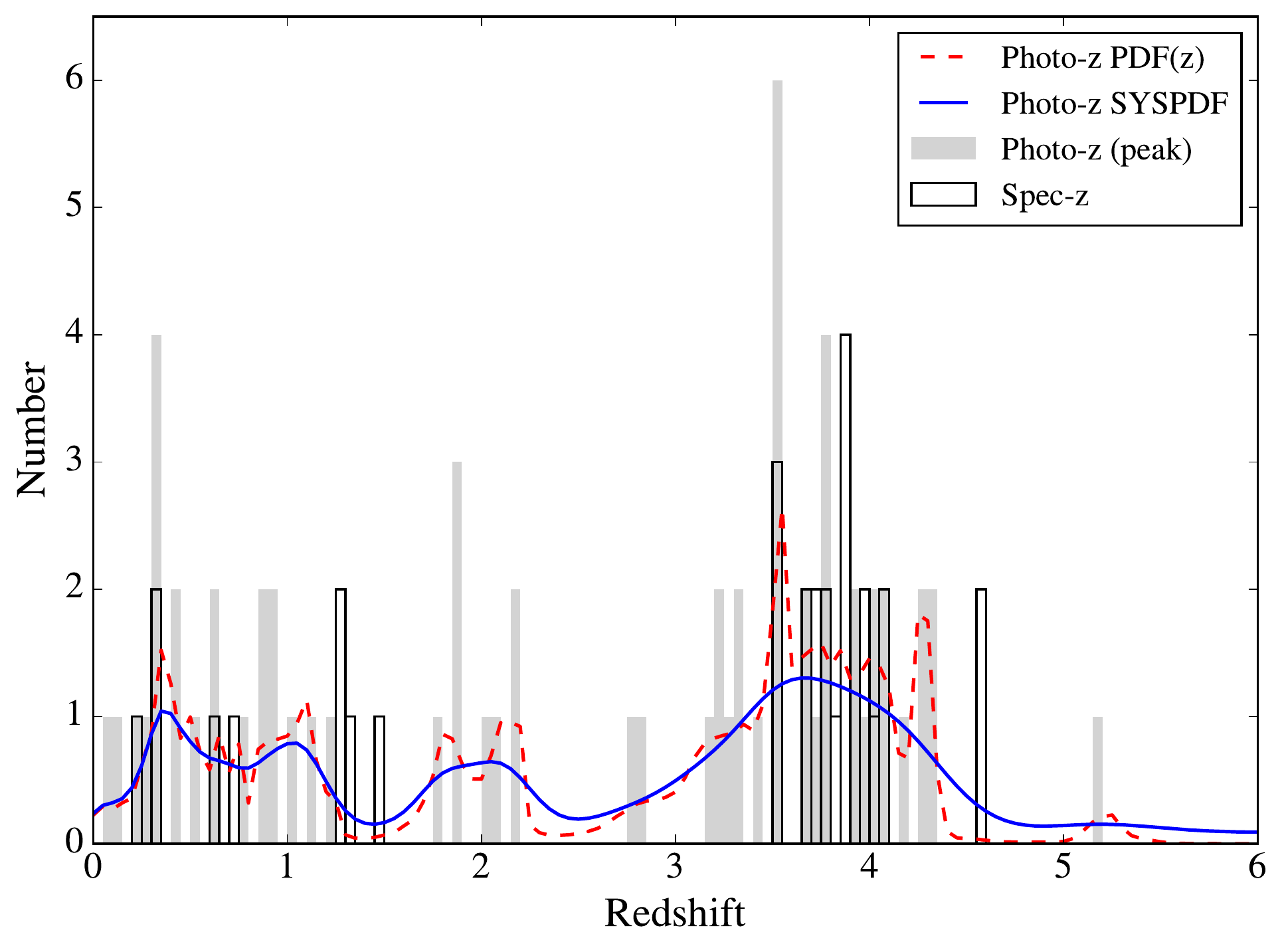}
    \end{tabular}
\caption{Redshift distribution for the 63 photo-z sources. The filled gray histogram presents the photo-z peak values, while the red dashed (blue solid) line shows the distribution when summing the PDF (SYSPDF) of all sources. For reference, we  plot the 30 dropouts with spectroscopic redshift (black histogram).} \label{redshiftHistSYS}
\end{figure}

In Fig.~\ref{RedshiftCompare2}, we show the comparison between the photo-z and spec-z. In addition to the most likely values (peak of the PDF, red crosses), we provide the full PDF(z). The grid is colour-coded with the relative redshift probability in each bin. We incorporate these systematic errors in the PDF(z) of our sample using the Eq.~\ref{eqsys}. In Fig.~\ref{outliers}, we show the individual corrected PDF(z) for two spectroscopically confirmed high-z sources, while in Fig.~\ref{redshiftHistSYS}, we show the redshift distribution of the dropout selected objects with spec-z and photo-z. In particular, the gray histogram represents the distribution of the peaks of the PDFz, while the blue solid line is the distribution when we sum the SYSPDF(z) of all sources together (corrected for $\tilde{\eta}$ and $\tilde{\sigma}$). For comparison, we show also the sum of the PDF(z) with red dashed line. In general, the two distributions (sum of PDF or SYSPDF and the simple source counts) agree with each other, showing that the majority of the sources have a PDF(z) with a single, narrow peak. In the same plot, we show the distribution of the dropouts with confirmed redshifts.


\subsection{Purity and completeness}\label{contamination}

In order to estimate the reliability and completeness of the Lyman Break technique, we used the spectroscopically confirmed high-z AGN sample. In the 4XXL-HSC area, we selected 28 confirmed high-z sources (Sect.~\ref{specz}). In Figs.~\ref{onoPlots} and \ref{akiyamaPlot}, we show the positions of spec-z sources in the colour-colour diagrams. The dropout selection criteria recovered 21/28 (75\%) sources. Out of the non-selected sources, two lie very close to the wedges, while one source lies inside the wedges but it did not pass the criteria concerning the signal-to-noise ratio. The spectra of the remaining sources show that strong emission lines fall between the windows of the photometric filters and affect the spectral colours.

\begin{figure}
\includegraphics[width=0.47\textwidth]{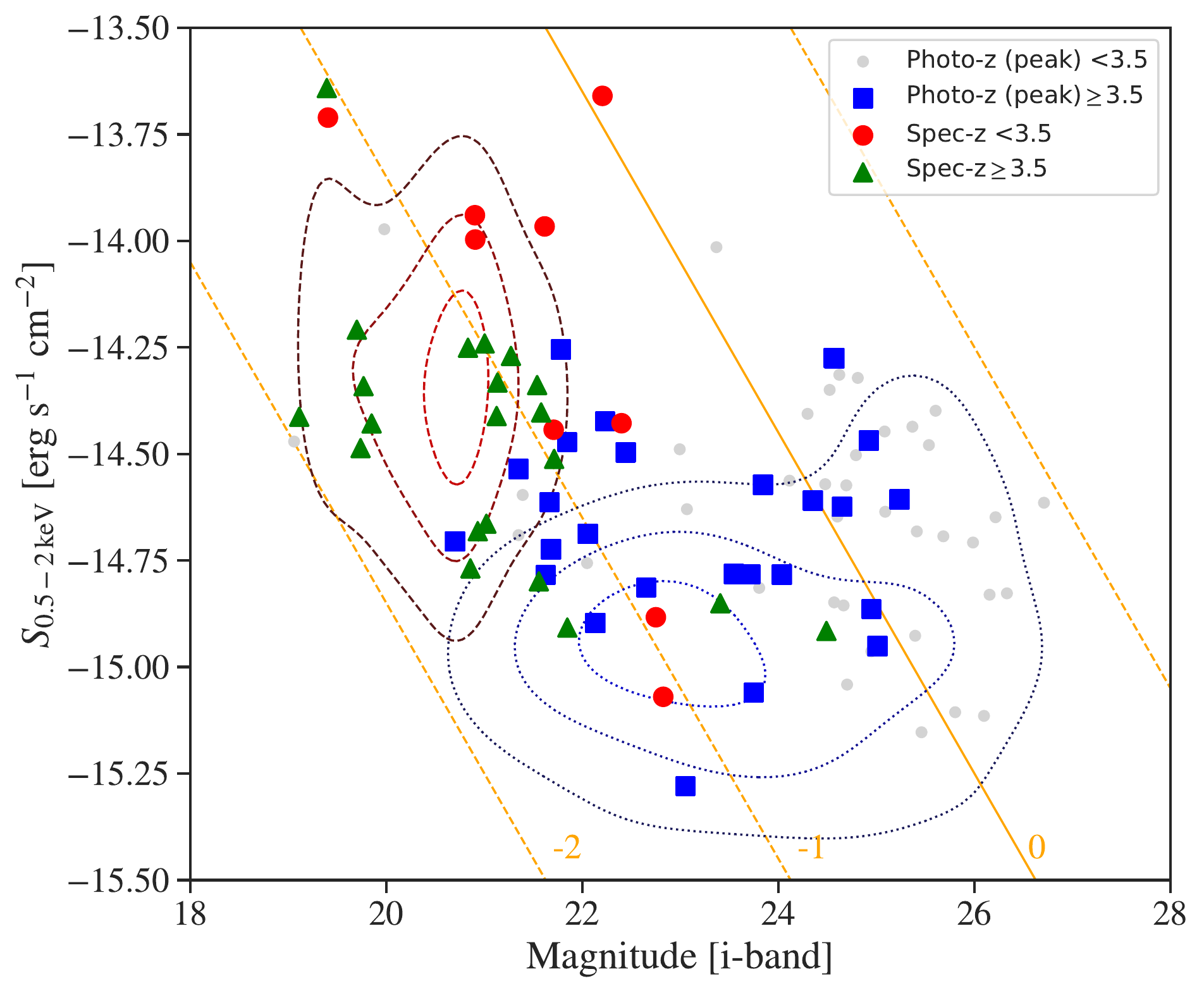}

\caption{Soft (0.5-2 keV) X-ray flux vs. optical (i-band) magnitude for the dropouts (gray points). Blue squares represent the sources with photo-z (peak) higher than 3.5. The green triangles (red circles) show the sources with $z_{\rm spec}\geqslant3.5$ ($z_{\rm spec}<3.5$). The solid line indicates the $\log(F_{\rm X}/F_{\rm opt})=0$ and the dashed lines from left to right correspond to $\log(F_{\rm X}/F_{\rm opt})=-2, -1, +1$, respectively. The X-ray fluxes are plotted in logarithmic scale and given in units of ergs cm\textsuperscript{-2} s\textsuperscript{-1}, while the optical magnitudes in the AB system. The dashed and dotted contours show the high-z sample of \citet{georgakakis2015} and \citet{marchesi2016}, respectively.}\label{fxfopt}
\end{figure}

Furthermore, the different colour-colour selection criteria allow contaminants, such as low-z galaxies and/or brown dwarfs, in the high-z sample. Concerning the stellar contamination, beside the X-ray emission that is a strong signature of AGN, we used the X-ray ($F_{\rm X}$) to optical flux ($F_{\rm opt}$) ratio, $F_{\rm X}/F_{\rm opt}$ \citep{maccacaro1988,barger2003,hornschemeier2003}. This relation is a well-known method to verify that a given object is an AGN and has been used in many studies \citep[e.g.][]{pouliasis2019}. The typical AGN population lies in the area between $\log(F_{\rm X}/F_{\rm opt})=\pm1$ with some spectroscopically confirmed AGN to be extended up to $\log(F_{\rm X}/F_{\rm opt})=\pm2$. Conversely, stars have low X-ray emission relative to their optical emission ($\log(F_{\rm X}/F_{\rm opt})\leqslant-2$) and thus they can be discerned very well from the AGN which are powerful X-ray emitters. In Fig.~\ref{fxfopt}, we plot the X-ray flux (0.5-2 keV) vs. the optical magnitude (i-band) of all the colour-colour selected sources having detections in the soft band. The sources with spectroscopic redshifts cover the area of the bright optically end, while the photo-z sources are fainter in the optical but with similar X-ray flux to the latter. 

The bulk of our high-z candidates are distributed over the whole area within $F_{\rm X}/F_{\rm opt}=\pm2$, that is indicative of their AGN nature. However, there are some rare cases of flaring ultra-cool dwarfs \citep{deluca2020} of type $\sim$M7-8 up to L1 with typical luminosity values of $\sim 10^{30}$ erg/s that could reach very high X-ray-to-optical ratios ($\log(L_{\rm X}/L_{\rm opt})\simeq-1$) and contaminate our sample. In our case though, all the SEDs of the final 63 photo-z sources are well fitted by AGN templates -- 98\% and 60\% of the sources have $\chi^2_{red}<6.5$ and $\chi^2_{red}<2$, respectively. Furthermore, almost $\sim$55\% of this sample is composed of extended sources. In Fig.~\ref{fxfopt}, we also plot the high-z AGN selected by \citet[XXL - smaller and shallower area than in our case,][]{georgakakis2015} and \citet[COSMOS - narrow and deep field,][]{marchesi2016} for reference. Our sources cover the regions of both these surveys. This is because, in our analysis, we used new X-ray observations in the XMM-XXL field in addition to the deep HSC data, we were able to push the X-ray and optical limits similar to COSMOS field.

Concerning the low redshift interlopers, \citet{ono2018} criteria include requirements in the detection threshold -- high signal-to-noise ratio for the redder bands and non detections in bluer bands -- to avoid as much as possible low-z contaminants. Even though applying these criteria, they estimated the fraction of low-z interlopers to be less than 10\% for faint (>24 mag) sources and about 40\% for brighter sources in the redshift range around z=4. \citet{akiyama2018}, studying only point-like sources, found that the expected contamination comes from compact objects at magnitudes >23 in the i band with a fraction of $\sim$30\%. In our study, we used the spectroscopic and photometric redshifts to estimate the possible contamination of the dropout selection method. Using the spectroscopic sample (31 sources), 68\% have z$\mathrm{_{spec}}>3.5$, while 32\% with lower redshifts contaminating our sample with an average redshift $z_{mean}=0.77$. For the remaining sources (63) lacking spectroscopic redshifts, we used the photo-z estimations. There are 37/63 ($\sim$58.7\%) with $z<3.5$. This is expected, if we take into account that at fainter magnitudes the colour selection becomes less reliable because of increased uncertainties.


\subsection{High-z sample -- summary}\label{contamination}

\begin{table}
\caption{Number of sources in different redshift bins.}              
\label{numbercounts}      
\centering                                      
\begin{tabular}{c c c c c c}          
\hline\hline                        
Redshift  & z$\mathrm{_{spec}}$ & z$\mathrm{_{phot}}$ & z$\mathrm{_{phot}}$  & Total \\
bins  &  & (peak) & (weighted) & number  \\
(1)  & (2) & (3) & (4) & (5)  \\

\hline                                   
   $z\geqslant3.5$ & 28 & 26 & 27.3 &  55.3  \\
    $z\geqslant4$  & 9 & 10 & 14.9  & 23.9  \\
    $z\geqslant5$  & 1 & 1 & 4.6   & 5.6 \\ 
\hline                                             
\end{tabular}
\tablefoot{(1): Redshift bins. (2): Number of spectroscopically confirmed sources. (3): Number of sources with photo-z (peak). (4): Effective number counts taking into account the SYSPDF(z) weights. (5): Total number of high-z sources in the soft 0.5-2 keV band using the spec-z and the sum of SYSPDF(z) contributing at this redshift bin.}
\end{table}

\longtab{
\begin{longtable}{c c c c c c c c c c c}
\caption{\label{speczcat} The sample of 91 sources (both spec-z and photo-z) used for the logN-logS determination.}\\
\hline\hline
XLSSU ID  &  Xra  &  Xdec  &  Ora  &  Odec  &  F\textsubscript{0.5-2 keV}   & F\textsubscript{2-10 keV}   &  z & z Ref. & $w_{i,3.5}$ & i band \\
(1) & (2) & (3) & (4) & (5) & (6) & (7) & (8) & (9) & (10) & (11) \\
\hline
\endfirsthead
\caption{continued.}\\
\hline\hline
XLSSU ID  &  Xra  &  Xdec  &  Ora  &  Odec  &  F\textsubscript{0.5-2 keV}   & F\textsubscript{2-10 keV}   &  z & z Ref. & $w_{i,3.5}$ & i band \\
(1) & (2) & (3) & (4) & (5) & (6) & (7) & (8) & (9) & (10) & (11) \\

\hline
\endhead
\hline
\endfoot
 J020150.3-070016 &30.4598&-7.0045&30.4596&-7.0036&5.6±1.5&-&4.080& 1,a &1.00&20.83\\
 J020222.9-051856 &30.5958&-5.3156&30.5967&-5.3154&3.7±1.2&-&4.10&0&0.89&22.23\\
 J020245.0-044223 &30.6877&-4.7066&30.6871&-4.7063&1.9±1.2&-&4.25&0&0.90&20.70\\
 J020253.8-065043 &30.7242&-6.8454&30.7240&-6.8457&3.2±1.6&-&3.860& 1,a &1.00&21.10\\
 J020300.2-070455 &30.7509&-7.0822&30.7502&-7.0816&2.3±1.2&-&3.40&0&0.48&23.07\\
 J020327.2-062106 &30.8636&-6.3519&30.8627&-6.3522&2.2±1.7&-&2.80&0&0.12&24.60\\
 J020353.1-050638 &30.9716&-5.1106&30.9710&-5.1097&4.6±1.8&-&3.715& 1,a &1.00&21.13\\
 J020423.8-051325 &31.0993&-5.2238&31.0993&-5.2232&1.7±1.2&-&3.768& 1,a &1.00&20.86\\
 J020431.3-053158 &31.1305&-5.5329&31.1300&-5.5327&3.6±3.3&17.6±13.1&1.20&0&0.28&25.37\\
 J020604.5-043108 &31.5188&-4.5191&31.5191&-4.5182&9.6±2.0&34.5±7.8&1.10&0&0.11&22.21\\
 J020703.8-070615 &31.7661&-7.1043&31.7657&-7.1037&5.5±1.5&-&4.30&0&0.89&21.78\\
 J020736.7-042540 &31.9032&-4.4280&31.9033&-4.4289&9.5±2.5&-&5.350& 1,a &1.00&22.28\\
 J020905.6-060049 &32.2735&-6.0138&32.2744&-6.0135&3.1±2.1&-&0.95&0&0.17&24.07\\
 J020922.0-044522 &32.3421&-4.7563&32.3410&-4.7577&1.6±1.3&-&4.00&0&0.89&23.55\\
 J020932.4-044407 &32.3853&-4.7355&32.3852&-4.7354&5.3±1.5&-&3.55&0&0.66&24.43\\
 J020946.8-062507 &32.4452&-6.4187&32.4428&-6.4193&3.9±1.8&-&1.05&0&0.11&23.39\\
 J020947.1-060417 &32.4466&-6.0715&32.4460&-6.0729&2.4±2.0&-&3.95&0&0.71&24.02\\
 J021039.4-055007 &32.6645&-5.8353&32.6648&-5.8351&2.0±1.3&-&0.75&0&0.11&21.35\\
 J021131.0-042126 &32.8792&-4.3575&32.8795&-4.3575&5.7±1.7&13.6±11.4&3.879& 1,a &1.00&21.00\\
 J021149.6-045007 &32.9567&-4.8353&32.9570&-4.8354&3.3±1.6&-&4.25&0&0.90&21.84\\
 J021157.5-060246 &32.9899&-6.0462&32.9896&-6.0469&1.9±0.8&-&0.10&0&0.26&25.99\\
 J021211.8-041056 &33.0494&-4.1823&33.0492&-4.1821&2.3±1.3&-&3.80&0&0.87&24.38\\
 J021338.5-051615 &33.4108&-5.2710&33.4097&-5.2710&8.9±0.8&26.5±7.4&4.540& 1,a &1.00&24.89\\
 J021344.6-052848 &33.4359&-5.4801&33.4363&-5.4798&1.4±1.1&-&3.20&0&0.21&24.43\\
 J021515.6-051915 &33.8153&-5.3210&33.8169&-5.3224&1.2±0.5&-&4.30&0&0.74&22.13\\
 J021527.2-060401 &33.8637&-6.0672&33.8637&-6.0667&5.3±2.4&14.9±13.6&4.065& 1,a &1.00&21.27\\
 J021544.0-045525 &33.9336&-4.9237&33.9334&-4.9228&2.1±0.8&-&3.522& 1,b &1.00&21.02\\
 J021613.8-040823 &34.0575&-4.1398&34.0578&-4.1409&3.9±0.9&-&3.522& 1,a &1.00&21.58\\
 J021646.8-041343 &34.1954&-4.2288&34.1951&-4.2296&2.6±0.9&-&3.20&0&0.19&24.33\\
 J021712.6-054108 &34.3029&-5.6857&34.3041&-5.6861&4.5±1.3&-&4.563& 1,a &1.00&21.54\\
 J021727.6-051718 &34.3651&-5.2885&34.3656&-5.2889&1.2±0.5&-&3.974& 1,b &1.00&21.85\\
 J021734.3-050513 &34.3930&-5.0872&34.3933&-5.0874&1.4±0.4&4.3±2.5&3.974& 1,b &1.00&23.41\\
 J021746.0-032951 &34.4419&-3.4977&34.4432&-3.4993&2.4±1.7&-&4.20&0&0.88&24.21\\
 J021747.0-054201 &34.4459&-5.7004&34.4469&-5.7002&1.4±0.7&7.1±7.6&0.20&0&0.19&26.15\\
 J021828.2-051551 &34.6175&-5.2642&34.6181&-5.2646&1.2±0.7&4.8±4.0&3.860& 1,b &1.00&23.87\\
 J021831.6-044358 &34.6321&-4.7330&34.6313&-4.7325&0.9±0.5&-&3.699& 1,b &1.00&23.77\\
 J021833.7-051713 &34.6405&-5.2871&34.6410&-5.2877&1.1±0.6&-&3.553& 1,b &1.00&23.33\\
 J021844.2-044826 &34.6843&-4.8073&34.6853&-4.8069&3.7±0.7&4.7±4.3&4.550& 1,a &1.00&19.85\\
 J021858.0-051933 &34.7417&-5.3260&34.7420&-5.3242&3.9±1.0&4.8±4.9&0.95&0&0.11&24.31\\
 J021915.4-042801 &34.8145&-4.4669&34.8146&-4.4680&1.5±0.5&-&3.761& 1,a &1.00&21.56\\
 J021952.4-034520 &34.9687&-3.7558&34.9689&-3.7570&1.6±1.4&21.2±13.6&5.20&0&0.90&23.72\\
 J021952.8-055958 &34.9703&-5.9995&34.9695&-5.9993&3.8±0.8&8.2±6.4&3.863& 1,a &1.00&19.11\\
 J022003.7-041201 &35.0154&-4.2003&35.0154&-4.2007&1.3±0.3&-&3.65&0&0.66&24.66\\
 J022007.1-042828 &35.0299&-4.4746&35.0297&-4.4748&1.5±0.5&3.9±2.6&0.90&0&0.11&23.54\\
 J022021.9-050427 &35.0914&-5.0744&35.0918&-5.0748&1.5±0.6&-&4.174& 1,b &1.00&20.83\\
 J022025.6-031819 &35.1068&-3.3054&35.1083&-3.3057&4.8±3.1&-&2.05&0&0.11&24.49\\
 J022026.6-024732 &35.1112&-2.7922&35.1110&-2.7923&2.6±1.6&-&0.35&0&0.12&24.54\\
 J022028.9-045402 &35.1208&-4.9008&35.1213&-4.9004&1.3±0.3&-&0.30&0&0.14&24.52\\
 J022033.9-040021 &35.1416&-4.0060&35.1417&-4.0055&1.5±0.7&-&3.55&0&0.69&22.65\\
 J022034.1-050700 &35.1423&-5.1168&35.1432&-5.1152&0.9±0.3&-&0.25&0&0.56&24.10\\
 J022037.0-034906 &35.1542&-3.8185&35.1538&-3.8176&2.0±1.1&14.7±9.9&2.10&0&0.11&25.46\\
 J022037.3-050045 &35.1556&-5.0126&35.1559&-5.0126&4.7±0.7&8.8±5.6&0.90&0&0.36&24.77\\
 J022100.5-042326 &35.2521&-4.3907&35.2522&-4.3908&4.5±0.5&8.9±3.3&3.710& 1,a &1.00&19.77\\
 J022126.3-055247 &35.3599&-5.8798&35.3603&-5.8789&2.7±1.2&-&1.85&0&0.14&23.82\\
 J022156.4-033339 &35.4852&-3.5610&35.4853&-3.5625&3.5±2.0&-&1.85&0&0.11&24.95\\
 J022156.6-055147 &35.4862&-5.8631&35.4858&-5.8636&3.8±1.2&-&3.847& 1,a &1.00&21.13\\
 J022214.3-041456 &35.5600&-4.2491&35.5594&-4.2490&2.4±0.5&5.7±2.4&4.00&0&0.89&21.67\\
 J022215.8-051603 &35.5662&-5.2676&35.5653&-5.2672&5.3±0.8&8.0±3.3&3.50&0&0.51&24.53\\
 J022240.3-050125 &35.6679&-5.0238&35.6685&-5.0238&1.1±0.4&-&3.90&0&0.84&24.92\\
 J022242.1-024652 &35.6757&-2.7812&35.6754&-2.7822&3.4±1.4&-&3.55&0&0.42&24.70\\
 J022247.4-045145 &35.6979&-4.8627&35.6972&-4.8623&0.5±0.2&-&3.65&0&0.78&23.05\\
 J022251.6-050713 &35.7154&-5.1203&35.7157&-5.1201&8.5±0.8&15.7±3.7&3.780& 1,a &1.00&19.74\\
 J022304.5-030124 &35.7689&-3.0235&35.7684&-3.0242&2.9±1.5&-&3.75&0&0.85&21.33\\
 J022307.0-041311 &35.7793&-4.2199&35.7802&-4.2195&1.6±0.5&-&3.75&0&0.56&24.03\\
 J022307.8-030839 &35.7829&-3.1443&35.7831&-3.1445&6.1±1.9&-&3.675& 1,a &1.00&19.70\\
 J022320.7-031824 &35.8365&-3.3068&35.8363&-3.3067&22.8±2.1&35.5±12.7&3.865& 1,a &1.00&19.39\\
 J022334.7-031237 &35.8949&-3.2103&35.8937&-3.2116&3.1±2.3&-&3.70&0&0.69&22.44\\
 J022351.1-043737 &35.9632&-4.6272&35.9634&-4.6276&2.5±0.5&-&0.40&0&0.11&20.69\\
 J022413.7-044012 &36.0574&-4.6701&36.0571&-4.6697&2.2±0.5&-&2.15&0&0.11&26.22\\
 J022444.1-043952 &36.1842&-4.6646&36.1838&-4.6648&1.6±0.5&-&3.50&0&0.60&21.62\\
 J022446.8-050522 &36.1951&-5.0896&36.1961&-5.0904&4.4±0.7&5.8±2.9&0.40&0&0.45&24.07\\
 J022448.7-052359 &36.2033&-5.4000&36.2037&-5.3995&1.7±0.4&-&0.65&0&0.11&21.85\\
 J022500.9-041557 &36.2538&-4.2659&36.2542&-4.2662&1.1±0.4&-&3.15&0&0.13&25.41\\
 J022505.3-053127 &36.2723&-5.5243&36.2725&-5.5238&0.8±0.4&-&3.55&0&0.69&23.74\\
 J022512.0-025709 &36.3002&-2.9527&36.3003&-2.9532&3.2±1.3&-&0.65&0&0.11&22.99\\
 J022520.4-051112 &36.3352&-5.1869&36.3354&-5.1870&0.7±0.3&6.8±2.3&2.90&0&0.13&26.09\\
 J022537.0-041009 &36.4043&-4.1694&36.4049&-4.1696&2.0±0.8&-&4.05&0&0.89&22.06\\
 J022612.2-053742 &36.5512&-5.6284&36.5507&-5.6288&2.3±0.9&-&3.35&0&0.34&25.09\\
 J022612.8-045400 &36.5534&-4.9001&36.5538&-4.9001&2.4±0.5&6.9±3.1&2.15&0&0.11&26.71\\
 J022614.5-053045 &36.5605&-5.5126&36.5598&-5.5127&1.8±0.6&-&3.90&0&0.89&21.68\\
 J022638.9-050120 &36.6622&-5.0223&36.6617&-5.0230&1.0±0.6&-&0.30&0&0.24&24.76\\
 J022709.5-044342 &36.7896&-4.7285&36.7908&-4.7278&2.0±0.4&6.2±2.7&3.35&0&0.40&25.06\\
 J022718.8-052809 &36.8283&-5.4694&36.8278&-5.4688&0.7±0.4&-&0.10&0&0.32& 25.46	\\
 J022801.0-035114 &37.0042&-3.8540&37.0022&-3.8522&0.7±0.5&-&1.85&0&0.11&25.80\\
 J022807.3-042758 &37.0306&-4.4662&37.0302&-4.4677&1.4±1.3&13.2±10.2&1.75&0&0.11&26.33\\
 J022931.5-044716 &37.3813&-4.7878&37.3818&-4.7863&3.3±2.3&-&0.50&0&0.11&19.06\\
 J023002.3-043118 &37.5098&-4.5219&37.5103&-4.5222&2.0±0.9&-&3.658& 1,a &1.00&20.93\\
 J023004.6-043418 &37.5196&-4.5719&37.5191&-4.5719&2.6±1.1&13.9±8.8&3.80&0&0.85&23.84\\
 J023058.6-041358 &37.7445&-4.2330&37.7445&-4.2328&3.0±2.3&-&4.015& 1,a &1.00&21.71\\
 J023223.4-045813 &38.0976&-4.9704&38.0971&-4.9702&3.3±1.3&-&3.30&0&0.26&24.84\\
 J023225.9-053729 &38.1083&-5.6248&38.1081&-5.6253&7.0±1.6&-&4.56& 1,a &1.00&22.77\\
\end{longtable}
\tablefoot{(1): Unique identifier in the 4XXL catalogue. (2,3): X-ray right ascension and declination (degrees). (4,5): Optical HSC right ascension and declination (degrees). (6): X-ray flux in the 0.5-2 keV band in $\rm 10^{-15}erg~s^{-1}~cm^{-2}$. (7): Flux in the 2-10 keV band in $\rm 10^{-15}erg~s^{-1}~cm^{-2}$. Infrared AGN luminosity. (8): Redshift. (9): Method used to compute z (0: photo-z and 1: spectroscopy), while the letter refers to redshift reference (a: \citet{ahumada2020}, b: \citet{hiroi2012} and c: \citet{coil2011,cool2013}. (10): The SYSPDF weight as defined in Eq.~\ref{wi} with $z_0=3.5$. Spec-z sources have $w_i=1$. (11): Magnitude in the HSC i filter (AB system). Furthermore, since we used the internal V4.2 X-ray catalogue in our analysis, the reader should be aware of that the ID's and the coordinates might be changed in the future V4.3 catalogue.}
}

\begin{figure}
   \begin{tabular}{c}
    \includegraphics[width=0.45\textwidth]{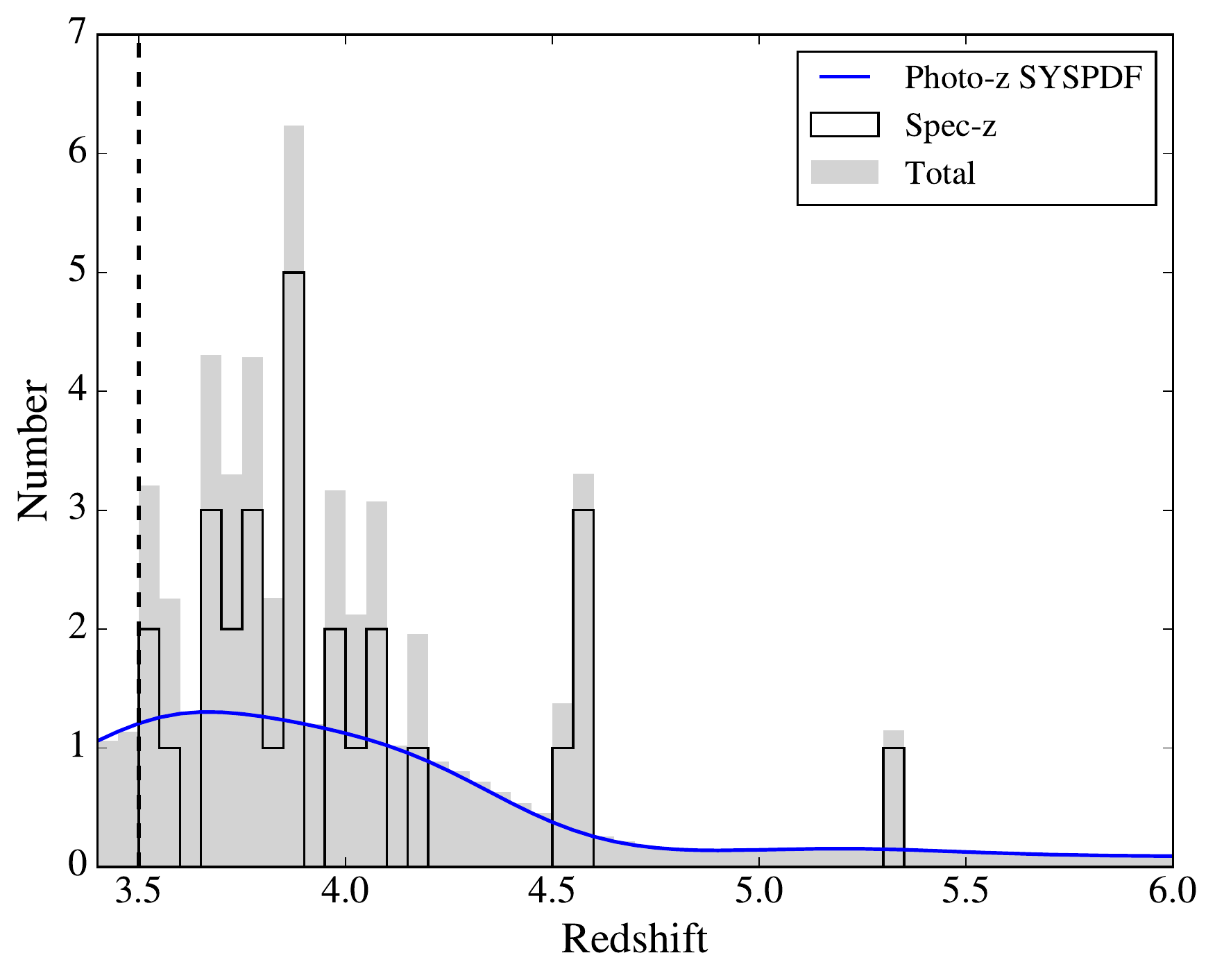}\\
    \end{tabular}
\caption{The redshift distribution of our final high-z sample (gray filled). We highlight the spec-z and the photo-z (sum of the SYSPDF). The vertical line indicate the redshift limit in our analysis ($z\geqslant3.5$).} \label{redshiftHist}
\end{figure}

Table~\ref{numbercounts} summarises the numbers of high-z sources in different redshift bins selected through our analysis. In particular, there are in total 28 sources with secure spectroscopic redshift at $z\geqslant3.5$ with 9 sources having $z\geqslant4$ and one source with redshift greater than five (ID= J020736.7-04254 with z$\mathrm{_{spec}}=5.35$). Furthermore, we selected 63 sources through colours that have not available spectra. Using the derived SYSPDF of the photometric redshifts, we were able to select additionally 26 sources with PDF peaks z$\mathrm{_{peak}}\geqslant3.5$. However, by considering the information containing in the full SYSPDF(z) of all the 63 sources, we were able to include in our analysis cases with lower probabilities being at high-z. The spec-z sources were assigned with weight equal to one. In this case, the effective number counts above redshift 3.5, 4 and 5 are 55.3, 23.9 and 5.6, respectively. Figure~\ref{redshiftHist} presents the redshift distribution of the final sources. This includes the spec-z sample (black histogram) and the sum of the SYSPDF(z) of all the 63 photo-z sources (blue line). Table~\ref{speczcat} lists all the 91 X-ray sources (28 spec-z and 63 photo-z) that we used for the logN-logS estimations in the next section.

We compared our final high-z sample with previous studies in the XMM-XXL northern field. Among the spectroscopic catalogue of \citet{menzel2016}, there are 55 $z\geqslant3.0$ sources used for the luminosity function calculation in \citet{georgakakis2015}. In their sample, there are 20 sources at $z\geqslant3.5$. Two sources with $z=3.67$ and $z=5.011$ do not have counterparts in the 4XMM-XXL catalogue. From the remaining, 15/18 sources were selected also through colours in our analysis. In our case, since we used the most updated X-ray observations that cover $\sim$40\% larger area and expanded our analysis to photo-z sources, we were able to select almost three times more high-z sources and almost 1.5 times when considering only spec-z sources since we used the most recent SDSS spectroscopy. Out of the 141 sources detected by \citet{vito2014} in various fields, 30 sources lie inside the SXDS field \citep{ueda2008} with 12 having $z\geqslant3.5$. Two X-ray sources with photometric redshift ($z=3.6$ and $z=4.09$) in their study do not have a counterpart in the 4XMM-XXL catalogue. Out of the remaining, we included all 8 sources with spec-z in our sample -- five of the them are dropouts. Finally, \citet{khorunzhev2019} compiled a catalogue of high-luminous, high-z sources ($\rm L_{X,2-10~keV} >10^{-15} erg~s^{-1}$) from the XMM-Newton serendipitous survey catalogue. Among these sources, 7 fall inside the 4XXL-N field and have a 4XXL counterpart. In our analysis, we included 5/7 sources with z$\mathrm{_{spec}}\geqslant3.5$.


\section{Number counts}\label{counts}

\begin{figure}
   \begin{tabular}{c}
    \includegraphics[width=0.45\textwidth]{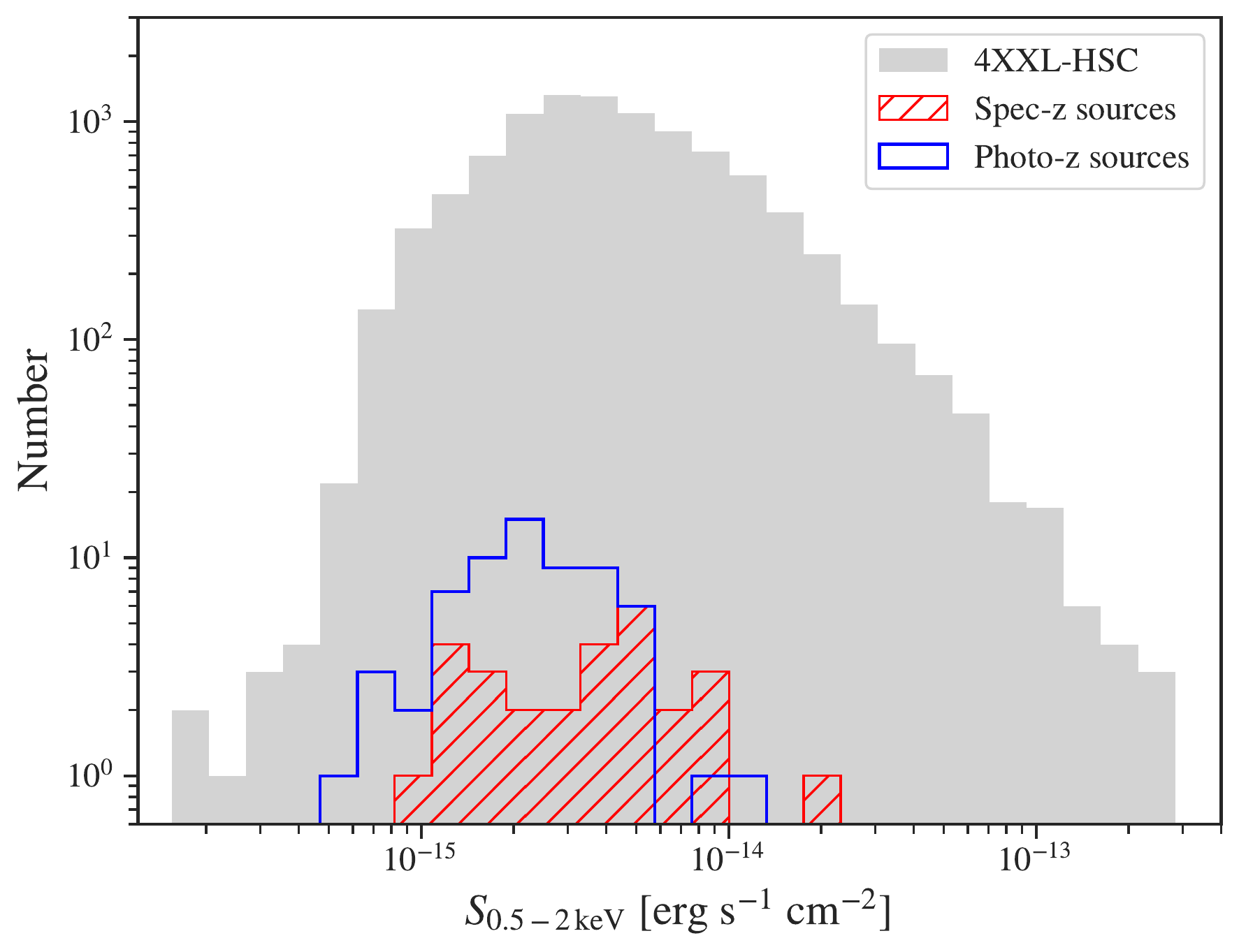} 
    \end{tabular}
\caption{Flux distribution in the soft (0.5-2 keV) band for all sources in the 4XXL-HSC catalogue (gray filled). The red dashed (blue solid) histogram represents the high-z sample with spectroscopic (photometric) redshift estimation.}\label{flux}
\end{figure}

\begin{figure}
    \includegraphics[width=0.47\textwidth]{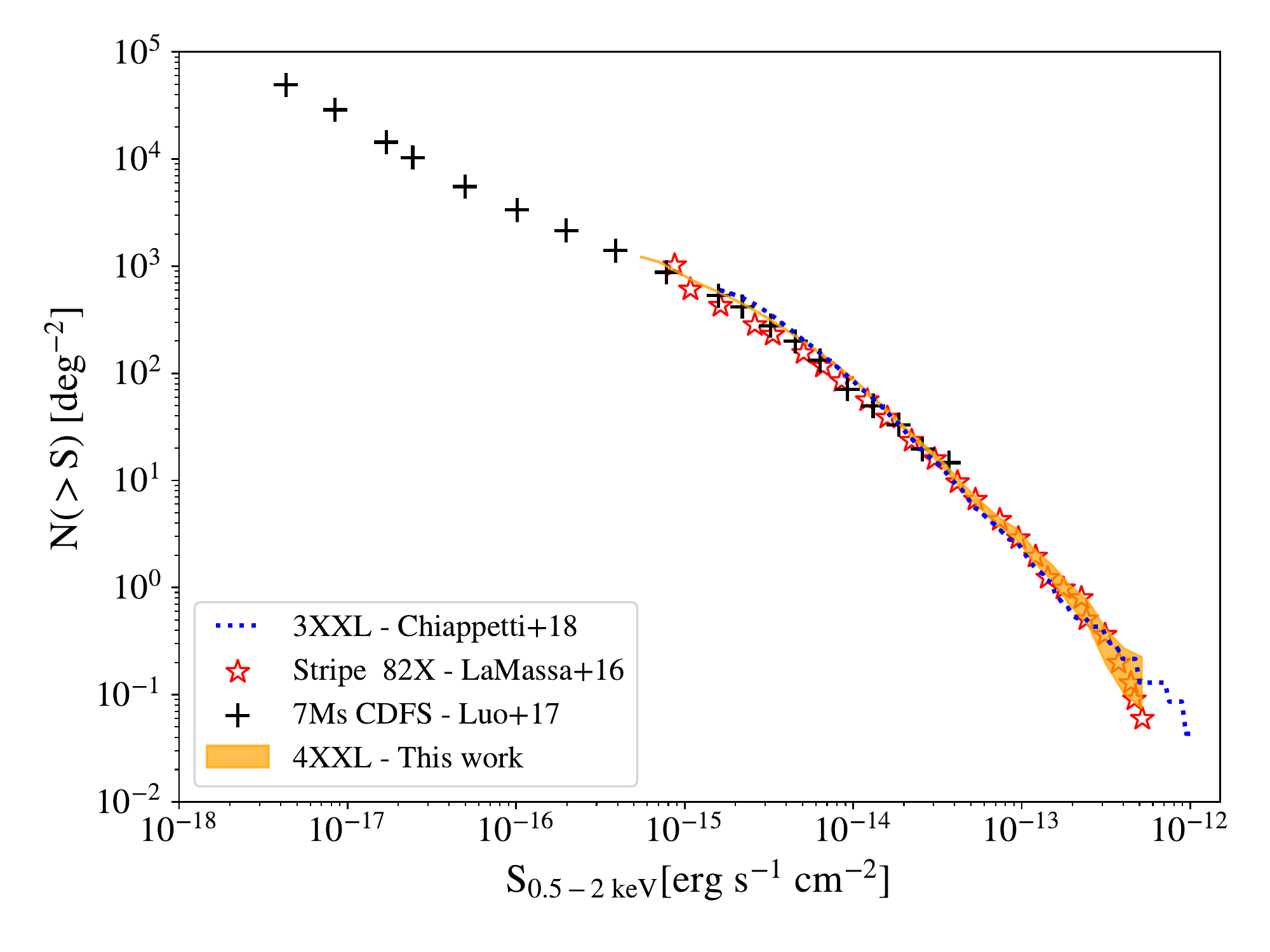}
\caption{The cumulative number counts for the whole 4XXL catalogue are presented with the shaded area highlighting the 1-$\sigma$ error in the soft 0.5-2 keV band. For reference, we over-plot the number counts derived by XXL Paper XXVII, \citet{luo2017} and \citet{lamassa2016}.}\label{integral}
\end{figure}

The cumulative number counts (the so-called logN-logS relation) is a very handful tool to describe and constrain the properties of the different AGN populations and test the theoretical assumptions concerning the evolution and the properties of the Universe. Taking into account the advantage of the large number of the high-z sources selected through our analysis, we were able to derive the number counts in the redshift bins $z\geqslant3.5$, $z\geqslant4$ and $z\geqslant5$ in the soft 0.5-2 keV X-ray band. In the redshift bin $z\geqslant6$, the main contribution in the number counts was originated from the scatter and the probability of being a catastrophic outlier components. Thus, we did not include this redshift bin in our results. In Fig.~\ref{flux}, we show the flux distribution in the soft 0.5-2 keV band for the 4XXL catalogue. We overplot as well the distribution of the high-z sample (spec-z and photo-z) that it ranges between $5 \times10^{-16}$ and $3 \times10^{-14} {\rm erg\,s^{-1} cm^{-2}}$, more than one order of magnitude. To calculate the integral form of the number count distribution we followed the traditional method defined as follows: 

\begin{ceqn}
\begin{equation}\label{wi}
N(>S_j)=\sum_{i=1}^{M}\frac{w_i}{\Omega_i}\mathrm{deg}^{-2}
\end{equation}
\end{ceqn}

where $N$ is the surface number density of sources with $S>S_j$, where $S_j$ is the lower edge of the bin. $\Omega_i$ is the solid angle in which a source with flux $S_i$ could have been detected and $M$ is the number of sources with $S>S_j$. We further included the weight, which is the integral of the SYSPDF(z) above a given redshift: 

\begin{ceqn}
\begin{equation}
w_i=\int_{z_0}^7 SYSPDF_i (z) dz
\end{equation}
\end{ceqn}

with $z_0$ the minimum redshift in each bin. $w_i$ is the probability that the object lies above redshift $z\geqslant z_0$ For spec-z sources we fixed this value to one.

\begin{figure}[h!]
   \begin{tabular}{c}
    \includegraphics[width=0.45\textwidth]{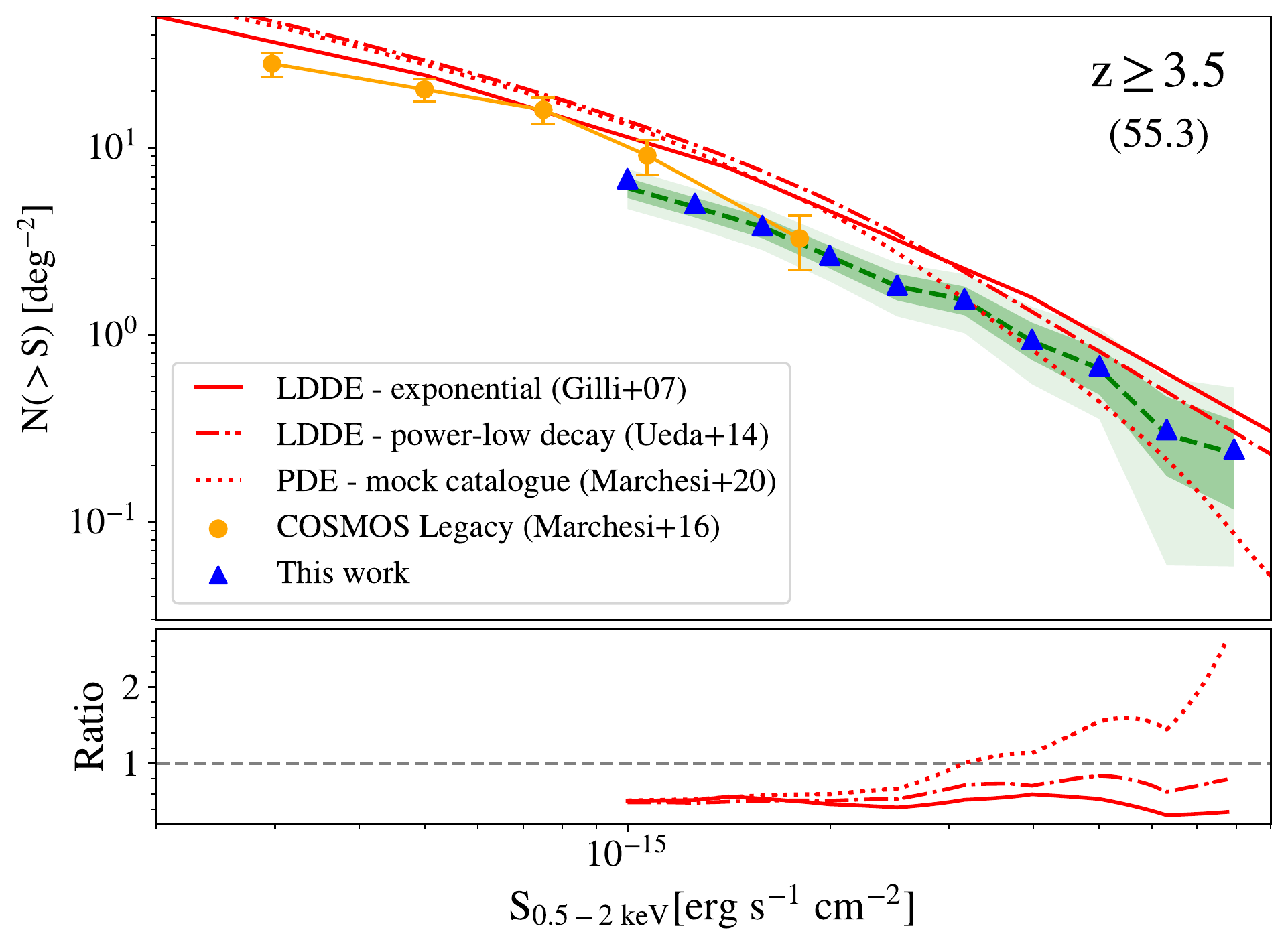}\\
    \includegraphics[width=0.45\textwidth]{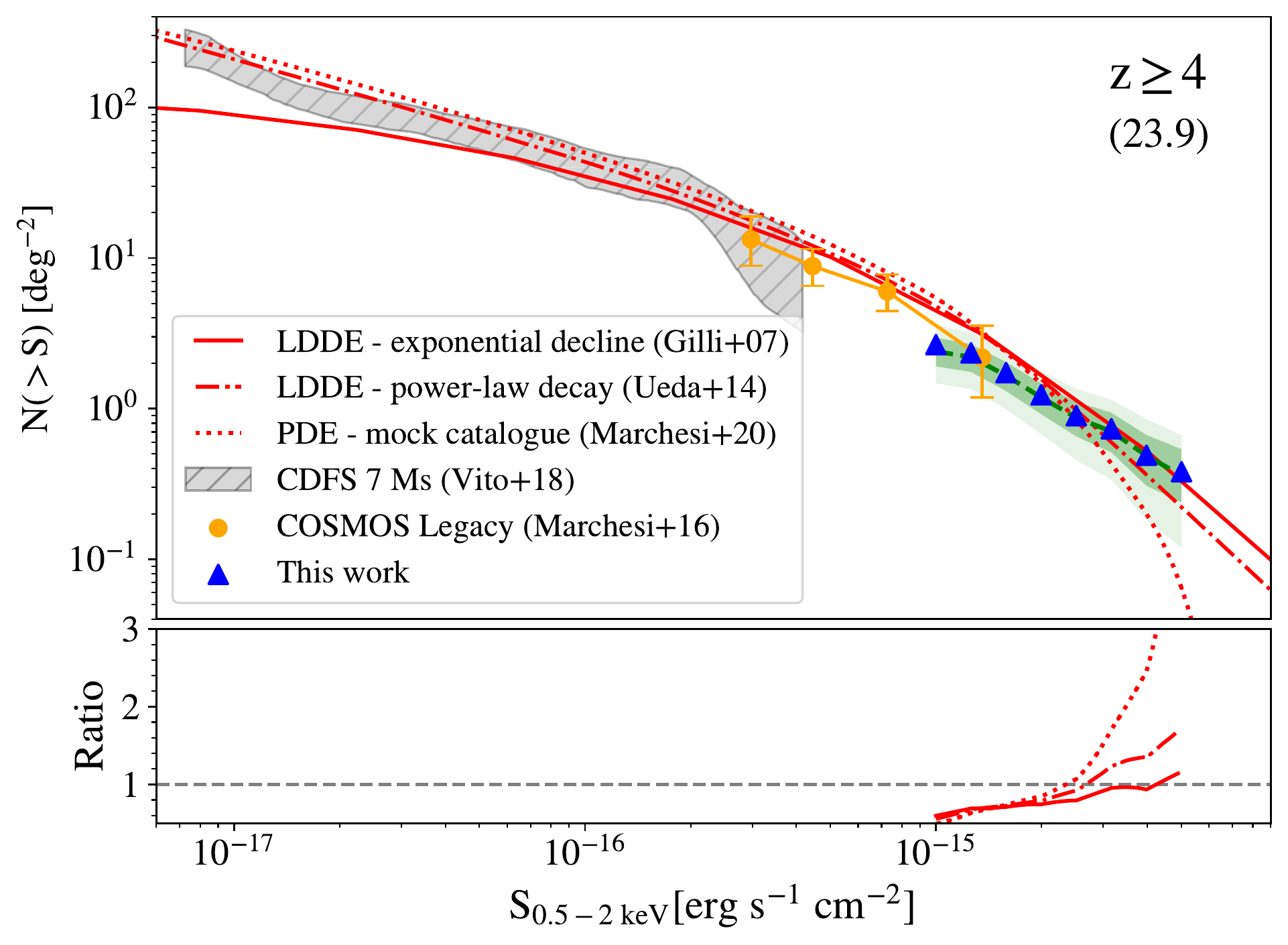}\\
    \includegraphics[width=0.45\textwidth]{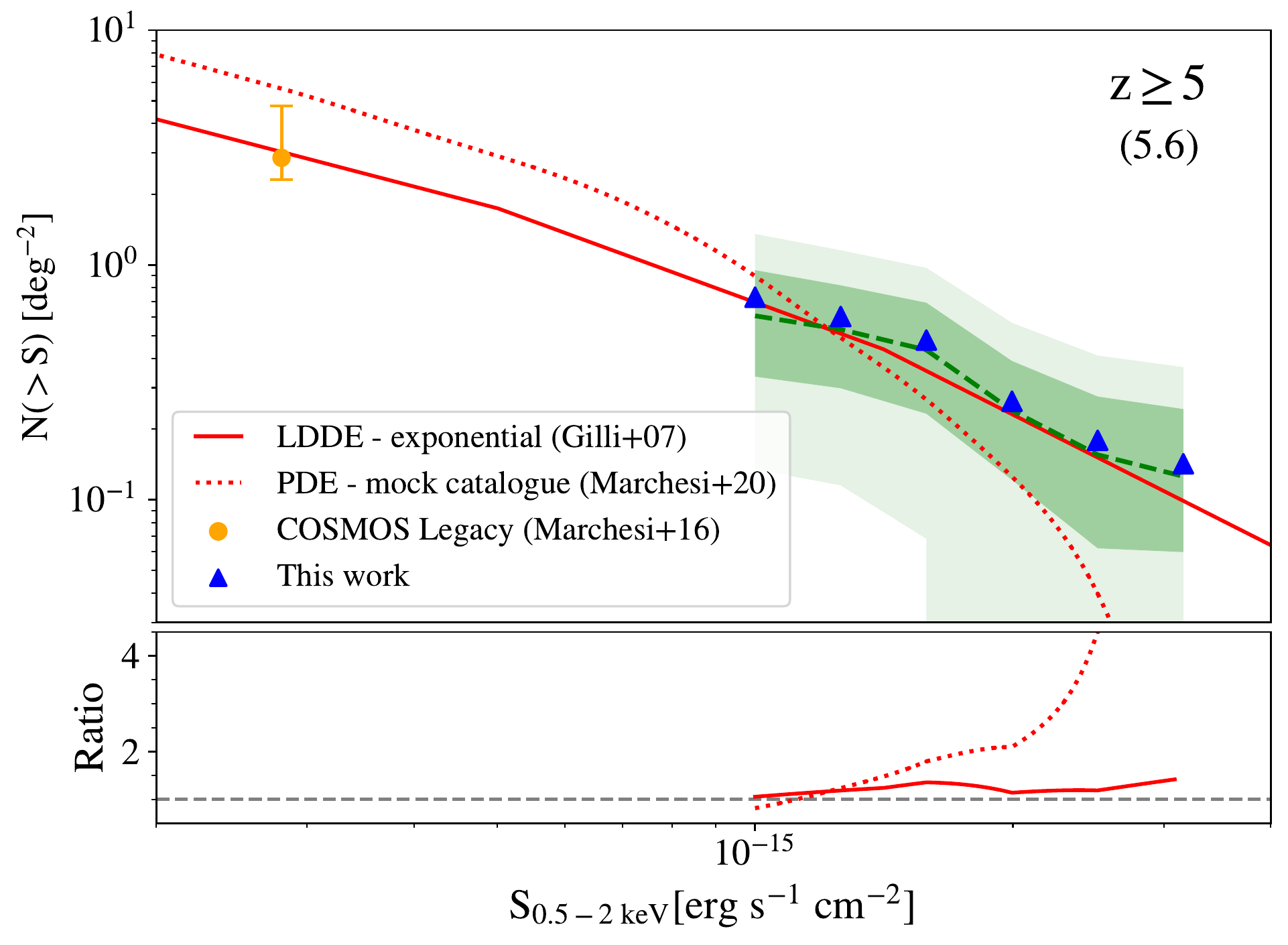}
    \end{tabular}
\caption{The source count distribution in the integral form corrected for the incompleteness due to optically selection effects for sources detected in the soft 0.5-2 keV band for the redshift bins $z\geqslant3.5$ (upper), $z\geqslant4$ (middle) and $z\geqslant5$ (lower). The light and dark shaded areas represent the 1$\sigma$ and 2$\sigma$ uncertainties as inferred from the bootstrap technique. The solid (dashed-dotted) line indicates the LDDE model predictions with an exponential decline (with a power-law decay) at high-z. The dotted line shows the mock catalogue based on the XLF by \citet{vito2014}. For reference, we show the data points derived by \citet{marchesi2016} and \citet{vito2018}. In the parenthesis, we give the effective number of sources in each redshift bin. Below each plot, we show the ratio between our data and the different models.}\label{logNlogS}
\end{figure}

We calculated the uncertainties using the bootstrap method. Thus, we generated randomly 10000 realizations of the high-z sample with the same size allowing for repetitions. Next, we assigned to each source a random redshift value following its SYSPDF(z). For sources with spec-z, we kept the spec-z value. At the end, for each list we calculated the logN-logS and the average, 1$\sigma$ and 2$\sigma$ values from all iterations. We used the area curve derived from the 4XXL catalogue in the soft band. As a sanity check, we calculated the logN-logS for the whole 4XXL catalogue. In Fig.~\ref{integral}, we show the cumulative numbers derived from this work compared to previous studies. In particular, we overplot the logN-logS trend derived in XXL Paper XXVII with the 3XXL catalogue, by \citet{luo2017} in the 7 Ms CDFS and by \citet{lamassa2016} in the Stripe 82X field. The number counts agrees very well with the aforementioned studies indicating that the area curve produced for the 4XXL sample is correct.

In order to use the area curve derived from the full 4XXL catalogue, we had to correct for the incompleteness due to the optically selection effects that appear in the 4XXL-HSC catalogue. In particular, as reported in Sec.~\ref{4xxlHSC} $\sim$88\% of the X-ray sources are matched with an optical counterpart. The incompleteness arises from the quality of the HSC data. The majority of non selected sources are affected by saturation, bad pixels, bright-object neighbouring or near-edge issues that we initially discarded from our sample. We may estimate the fraction of the X-ray sources that have a good HSC match for each of the logN-logS bins and correct for this incompleteness by adding an additional weight in Eq.~\ref{wi}. Furthermore, the 4XXL-HSC area is much smaller compared to the total area covered by the 4XXL data, because the HSC data do not cover the full area of 4XXL and also because we excluded the HSC masked areas due to bright stars. We corrected this by normalising the number counts to the total area. Finally, we examined if the dropout sources excluded from our sample due to their high reduced $\chi ^2$ values could affect the number counts. Out of seven sources, one is spectroscopically confirmed at $z_{spec}=1.24$ and one is found in the outskirts of the nearby galaxy 2MASX J02210771-0459574 ($z=0.13$) with biased photometry. Concerning the remaining five sources, we re-calculated the number counts in the three redshift bins assuming a flat SYSPDF(z) over the whole redshift range ($z=0-7$) for these sources. Analysing the derived logN-logS, we found no more than 1\%, 2\% and 4\% difference in the redshift bins $z\geqslant3.5$, $z\geqslant4$ and $z\geqslant5$, respectively, and only for the two to three faintest bins in each case.

In Fig.~\ref{logNlogS}, we show the cumulative source distributions in the different redshift bins corrected for the incompleteness due to the optically selection function and the HSC coverage. We compare our results to the predictions of the X-ray background synthesis model by \citet{gilli2007}. This model is based on the optical luminosity function parametrised with a luminosity-dependent density evolution (LDDE) model and an exponential decline at high-z (solid line). We show also the number counts of the mock catalogue of X-ray selected AGN generated by \citet[dotted line,][]{marchesi2020}. This catalogue is based on the X-ray luminosity function (XLF) by \citet{vito2014}, which assumes a pure density evolution \citep[PDE,][]{schmidt1968}. Finally, we compare the number counts with the \citet{ueda2014} model, which is composed of a LDDE model similar to \citet{gilli2007} model but instead of an exponential decline there is, additionally, a power-law decay. The \citet{ueda2014} model was built with a much larger sample of AGN compared to \citet{gilli2007} model over the redshift range from 0 to 5. Since, the upper redshift limit is five, we only show this model (dashed-dotted line) in the first two redshift bins. Furthermore, we compare our results in the faint end with the number counts derived by \citet{marchesi2016}. They used a high-z sample from the COSMOS Legacy survey. For the $z\geqslant4$ bin, we also include the data from \citet{vito2018} in the 7 Ms Chandra Deep Field-South and 2 Ms Chandra Deep Field-North at even fainter fluxes. Finally, to assist the plot interpretation and highlight the comparison of our number counts to the various models, we plot the ratio between them. The horizontal dashed lines show ratios equal to one.

In the redshift bin $z\geqslant3.5$, our number counts agree with the results of \citet{marchesi2016} in the bright end of their flux distribution. However, at lower fluxes ($\rm \sim10^{-15} erg~s^{-1}~cm^{-2}$) our results suggest lower number counts. This difference ($\sim$25\%) may arise due to the fact that \citet{marchesi2016} included all the X-ray selected sources in the field whose PDF does contain significant probability at $z\geqslant3.5$. We derived the photo-z only for the dropout candidates and not the full 4XXL catalogue. Also, our sample is not corrected for selection effects that are difficult to estimate. Such biases include the incompleteness caused by missed sources, either sources with no spectroscopic information or sources missed by the dropout method due to the indistinct borders of the selection colour criteria. Alternatively, the difference in the number counts between our results and those from \citet{marchesi2016} could be due to the cosmic variance. As pointed out in XXL Paper XXVII, the number counts in the COSMOS field ($\sim$2\,deg$^2$) are slightly overestimated. In order to minimise the cosmic variance, the bright end of the count distribution requires areas larger than $\sim$5-10\,deg$^2$ \citep{civano2016}. At fluxes higher than $\rm \sim3\times10^{-15} erg~s^{-1}~cm^{-2}$, our analysis agrees with the \citet{vito2014} or the \citet{ueda2014} models within 1$\sigma$ and with the \citet{gilli2007} model within 2$\sigma$. At the faint end, our data points seem to underestimate the number counts compared to all models by a factor of 2. This difference in our estimations could be due to the incompleteness of the dropout selection criteria. \citet{akiyama2018} showed that a fraction of X-ray AGN do not follow the dropout selection criteria. This concerns both blue and red sources, and the incompleteness could reach 20\%. Even though we updated the selection criteria (Sect.~\ref{4.1}) and we have recovered a portion of the red sources, we miss those with bluer colours.

In the redshift bin $z\geqslant4$ and $z\geqslant5$, our results are in good agreement with the COSMOS Legacy data points and also with the model predictions of \citet{gilli2007} within the uncertainties (1$\sigma$). Furthermore, it is the first time that we derive the number counts in the redshift bin $z\geqslant5$ at these bright fluxes ($\rm f_{0.5-2~keV} > 10^{-15} erg~s^{-1}~cm^{-2}$). Previously, \citet{marchesi2016} obtained the logN-logS for the same redshift bin, but at fainter fluxes. The \citet{vito2014} model, even though it agrees well with the data of \citet{vito2018}, underestimates the number counts towards bright fluxes ($\rm f_{0.5-2~keV} > \sim5 \times10^{-15} erg~s^{-1}~cm^{-2}$). This is because this model is based on small-area surveys, thus the bright end of the XLF is poorly sampled. For a better understanding of these discrepancies a joint analysis of shallow and deep surveys, using consistent methods, is required. Then, the computation of the high-z XLF should be less biased over the full flux distribution.


\section{Conclusions}\label{summary}
In this work, we selected an X-ray sample of high-z sources in the XMM-XXL northern field. We used the most updated X-ray observations in combination with the deep optical HSC data. In particular, we selected all the spectroscopically confirmed AGN and complemented this sample with high-z candidates using the Lyman Break technique. To verify the latter, we derived the photometric redshifts using \texttt{X-CIGALE}, a SED fitting algorithm. Having a large sample of high-z sources, we were able to put strong constraints in the number counts for different redshift intervals in the bright end of the flux distribution ($\rm f_{0.5-2~keV} > 10^{-15} erg~s^{-1}~cm^{-2}$). Our main results can be summarised as follows:

\begin{enumerate}

\item We applied the colour-selection criteria as defined in \citet{ono2018} and \citet{akiyama2018} and selected in total 101 high-z candidates, both point-like and extended sources. Moreover, we identified 28 high-z ($z\geqslant3.5$) sources using different spectroscopic catalogues available in the 4XXL area.

\item The photometric redshifts of the dropouts were obtained using the \texttt{X-CIGALE} algorithm. We calculated the performance of this method using different statistical approaches and resulted in small scatter $\rm \tilde{\sigma}=0.04^{+0.05}_{-0.02}$ and a fraction of outliers $\rm \tilde{\eta}=19.3^{+23.2}_{-14.4}\%$ with less than 10\% when focusing on the $z\geqslant3.5$ area. 

\item We estimated the possible contamination of the Lyman Break technique by stellar objects and low-z interlopers. In addition to the X-ray emission, the SEDs of our candidates were well fitted with AGN templates. Using additionally the $F_{\rm X}/F_{\rm opt}$ relation, we were certain that our sample is not contaminated by brown dwarfs.

\item For the low-z interlopers, we used the 30 sources with available spec-z. We found that 35\% of the colour-selected sources are at low redshifts. The percentage for the photo-z sample is higher but this is due to sources with fainter magnitudes and higher uncertainties and thus allowing red galaxies to enter the selection criteria wedges.

\item At the end, we were able to select 54 high-z sources (28 zspec). Our sample is three times (1.5 times considering only spec-z sources) larger than previous studies in the field. Additionally to these sources, for the logN-logS estimation we used also the dropout sources that have z$\mathrm{_{peak}}<3.5$ but the contribution of their SYSPDF above $z\geqslant3.5$ is not negligible.

\item Taking the advantage of out high-z sample, we were able to constrain with high accuracy the logN-logS relation in the redshift bins $z\geqslant3.5$, $z\geqslant4.0$ and, for the first time, $z\geqslant5$ at relatively bright fluxes ($\rm f_{0.5-2~keV} > 10^{-15} erg~s^{-1}~cm^{-2}$) that was previously poorly constraint. Our analysis agrees with the LDDE model predictions similar to the optical wavelengths. Compared to previous studies, there were some discrepancies that are caused by unavoidable incompleteness of our sample or due to the cosmic variance between pencil beam and large area surveys.

\end{enumerate}

We conclude that the combination of large-area X-ray surveys, such as XMM-XXL, with deep optical photometry is essential to identify with high completeness the AGN population in the early Universe. Wide X-ray surveys allow rare AGN to be found, while the deep optical data with lower uncertainties may contribute to their location and redshift estimations. The latter are critical for constraining the AGN sky density across the Universe and studying the evolutionary models of the SMBHs and their effect on their host galaxy environment.

\begin{acknowledgements}\label{ackn}
The authors are grateful to the anonymous referee for a careful reading and helpful feedback. We acknowledge Dr. Stefano Marchesi for computing and providing the COSMOS legacy logN-logS data points at $z\geqslant3.5$. EP and IG acknowledge financial support by the European Union's Horizon 2020 programme "XMM2ATHENA" under grant agreement No 101004168. The research leading to these results has received funding from the European Union's Horizon 2020 Programme under the AHEAD2020 project (grant agreement n. 871158). RG acknowledges financial contribution from the agreement ASI-INAF n. 2017-14-H.O. The Saclay team acknowledges long-term support from the Centre National d'Etudes Spatiales".

XXL is an international project based around an XMM Very Large Programme surveying two 25 deg2 extragalactic fields at a depth of $\rm \sim 6 \times 10^{-15} erg~s^{-1}~cm^{-2}$ in the [0.5-2] keV band for point-like sources. The XXL website is http://irfu.cea.fr/xxl. Multi-band information and spectroscopic follow-up of the X-ray sources are obtained through a number of survey programmes, summarised at http://xxlmultiwave.pbworks.com/.
This research has made use of the SIMBAD database \citep{wenger2000}, operated at CDS, Strasbourg, France and, also, of NASA's Astrophysics Data System. 
This research made use of Astropy, a community-developed core Python package for Astronomy (Astropy Collaboration et al. 2013 \url{http://www.astropy.org}). This publication made use of TOPCAT \citep{taylor2005} for table manipulations. 
The plots in this publication were produced using Matplotlib, a Python library for publication quality graphics \citep{hunter2007}.
Based on observations obtained with MegaPrime/MegaCam, a joint project of CFHT and CEA/DAPNIA, at the Canada-France-Hawaii Telescope (CFHT) which is operated by the National Research Council (NRC) of Canada, the Institut National des Sciences de l'Univers of the Centre National de la Recherche Scientifique (CNRS) of France, and the University of Hawaii. This work is based in part on data products produced at Terapix and the Canadian Astronomy Data Centre as part of the Canada-France-Hawaii Telescope Legacy Survey, a collaborative project of NRC and CNRS. 
The Hyper Suprime-Cam (HSC) collaboration includes the astronomical communities of Japan and Taiwan, and Princeton University. The HSC instrumentation and software were developed by the National Astronomical Observatory of Japan (NAOJ), the Kavli Institute for the Physics and Mathematics of the Universe (Kavli IPMU), the University of Tokyo, the High Energy Accelerator Research Organization (KEK), the Academia Sinica Institute for Astronomy and Astrophysics in Taiwan (ASIAA), and Princeton University. Funding was contributed by the FIRST program from the Japanese Cabinet Office, the Ministry of Education, Culture, Sports, Science and Technology (MEXT), the Japan Society for the Promotion of Science (JSPS), Japan Science and Technology Agency (JST), the Toray Science Foundation, NAOJ, Kavli IPMU, KEK, ASIAA, and Princeton University. This paper makes use of software developed for the Large Synoptic Survey Telescope. We thank the LSST Project for making their code available as free software at  http://dm.lsst.org. This paper is based [in part] on data collected at the Subaru Telescope and retrieved from the HSC data archive system, which is operated by Subaru Telescope and Astronomy Data Center (ADC) at National Astronomical Observatory of Japan. Data analysis was in part carried out with the cooperation of Center for Computational Astrophysics (CfCA), National Astronomical Observatory of Japan. 

Funding for the Sloan Digital Sky 
Survey IV has been provided by the 
Alfred P. Sloan Foundation, the U.S. 
Department of Energy Office of 
Science, and the Participating 
Institutions. 
SDSS-IV acknowledges support and 
resources from the Center for High 
Performance Computing  at the 
University of Utah. The SDSS 
website is www.sdss.org.
SDSS-IV is managed by the 
Astrophysical Research Consortium 
for the Participating Institutions 
of the SDSS Collaboration including 
the Brazilian Participation Group, 
the Carnegie Institution for Science, 
Carnegie Mellon University, Center for 
Astrophysics | Harvard \& 
Smithsonian, the Chilean Participation 
Group, the French Participation Group, 
Instituto de Astrof\'isica de 
Canarias, The Johns Hopkins 
University, Kavli Institute for the 
Physics and Mathematics of the 
Universe (IPMU) / University of 
Tokyo, the Korean Participation Group, 
Lawrence Berkeley National Laboratory, 
Leibniz Institut f\"ur Astrophysik 
Potsdam (AIP),  Max-Planck-Institut 
f\"ur Astronomie (MPIA Heidelberg), 
Max-Planck-Institut f\"ur 
Astrophysik (MPA Garching), 
Max-Planck-Institut f\"ur 
Extraterrestrische Physik (MPE), 
National Astronomical Observatories of 
China, New Mexico State University, 
New York University, University of 
Notre Dame, Observat\'ario 
Nacional / MCTI, The Ohio State 
University, Pennsylvania State 
University, Shanghai 
Astronomical Observatory, United 
Kingdom Participation Group, 
Universidad Nacional Aut\'onoma 
de M\'exico, University of Arizona, 
University of Colorado Boulder, 
University of Oxford, University of 
Portsmouth, University of Utah, 
University of Virginia, University 
of Washington, University of 
Wisconsin, Vanderbilt University, 
and Yale University.
\end{acknowledgements}

\bibliographystyle{aa}

\bibliography{bibfile} 

\end{document}